# Sustained Orbital Motion Driven by Circularly Polarized Light in Nanoscale Stator–Rotor Architectures

Shiye Du [a, b, c, †], Juanshu Wu [a, b, c, †], Xin Chen [e], Mouhong Lin [d, *], Hongyu Chen [a, b, c, *]

[a] Department of Chemistry, School of Science, Westlake University, Hangzhou 310030, China

[b] Research Center for Industries of the Future, Westlake University, Hangzhou 310030, China

[c] Institute of Natural Sciences, Westlake Institute for Advanced Study, Hangzhou 310024, China

[d] New Cornerstone Science Laboratory, State Key Laboratory of Extreme Photonics and Instrumentation, College of Optical Science and Engineering, Zhejiang University, Hangzhou, China

[e] Suzhou Laboratory, Suzhou 215123, China

[†] These authors contributed equally to this work.

[*] All correspondence should be addressed to linmouhong@zju.edu.cn and chenhongyu@westlake.edu.cn

## Abstract

Light-induced forces and torques offer a versatile strategy for remotely actuating microscopic objects, enabling breakthroughs including optical trapping and light-driven nanomachines. Extending such optical actuation to sustained cyclic motion, however, requires fundamentally distinct designs and, in particular, forces capable of delivering nonzero mechanical work over repeated cycles. Here, we demonstrate *via* simulations a nanoscale stator–rotor architecture, which achieves persistent orbital motion of a spherical rotor under circularly polarized plane-wave illumination. An off-center rotor displacement defines a stator–rotor structural polarity. Optical helicity coupled to this polarity generates a tangential force perpendicular to the instantaneous stator–rotor direction; as the rotor moves around the stator, the force direction rotates with it. Electrodynamic symmetry analysis identifies this response as arising from antisymmetric, nonconservative forces that produce nonzero closed-cycle work, whereas linear polarization yields symmetric, conservative responses with zero closed-cycle work. Material screening and Bayesian optimization identify an Ag-nanoshell design region, while Pareto analysis balances tangential actuation and radial confinement. A balanced candidate is predicted to lower the maximum steady-state temperature rise from 56.2 to 20.9 K relative to Au at the same light intensity. Brownian-dynamics simulations further show confined, helicity-defined orbital motion, with all 32 trajectories retaining the same circulation direction. These results establish a symmetry-guided route toward sustained cyclic optical actuation at the nanoscale.

## Graphic TOC

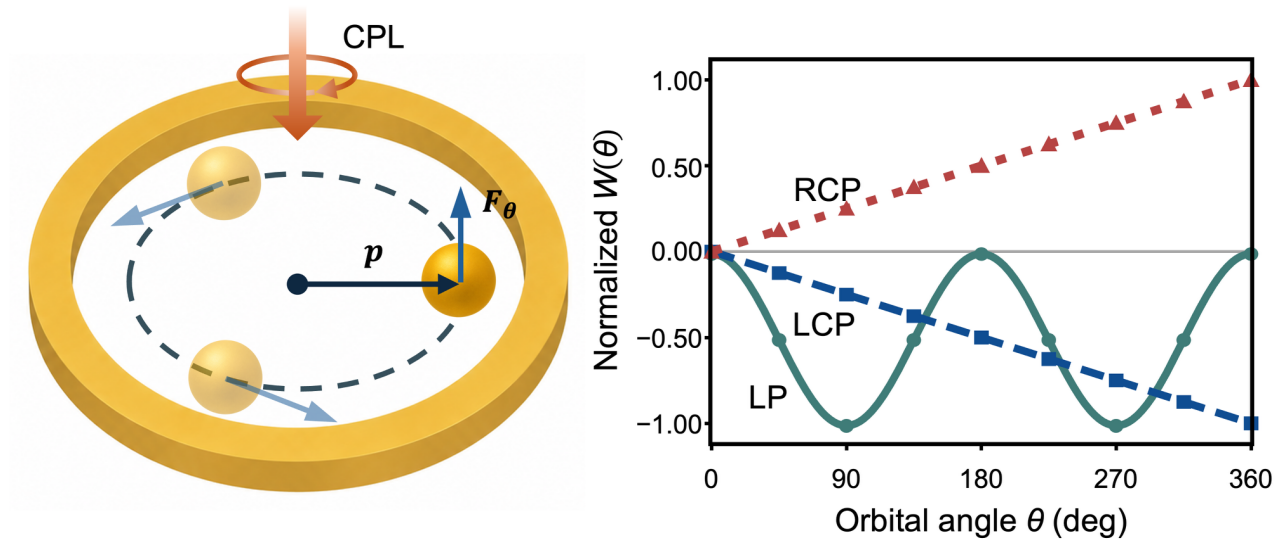

Light-triggered motions have long garnered interest spanning fundamental scientific research and futuristic conceptual imagination, fueling innovative developments in nanoscience and space exploration alike.[1-4] The underlying principle centers on remote, contactless manipulation *via* light-induced forces and torques: light transfers energy and momentum to physical entities, from minuscule nanoparticles all the way to large distant spacecraft.[3-5]

Optical forces nowadays underpin a broad spectrum of applications, including optical trapping,[2, 6] microfluidic transport,[7] colloidal assembly,[8] and light-driven nanomachines.[9, 10] Their mechanical effects are of particular importance at microscopic length scales. Owing to the ultralow mass and inertia of tiny objects, photon momentum can readily induce substantial mechanical motion. Accordingly, optical actuation has emerged as a powerful strategy for constructing remotely actuated, programmable microscopic machinery.[6, 10, 11]

Progress in this field has enabled increasingly diverse forms of microscopic motions. In optical tweezers, focused light fields generate gradient forces that trap particles and move them by steering the focal position.[2, 12] Conventional radiation pressure associated with absorption and scattering primarily drives motion along the optical-momentum direction.[1, 6] The spin and orbital angular momentum of light can further induce alignment, self-spinning, and rotation of particles with appropriate optical responses.[5, 13-16] More recently, structured illumination,[17-20] nearby interfaces,[21, 22] multipolar interference,[23] and asymmetric nanostructures[24, 25] have enabled transverse optical forces perpendicular to the incident momentum,[26, 27] broadening the accessible motion to include micro-objects rotation,[9] integrated micromachines steering[10, 24], and collective rotational and nonequilibrium dynamics[28-30].

Despite these advances, realizing sustained orbital motion remains challenging, particularly for an achiral, shape-isotropic rotor under normally incident, spatially uniform plane-wave illumination. Closed-path rotational motion is a fundamental mechanical motif in both engineered and biological systems, ranging from macroscopic motors to molecular rotary machines such as ATP synthases[31, 32] and bacterial flagella[33, 34].

The bacterial flagellar motor is arguably the most sophisticated rotary nanomachine in nature. Ion-conducting stator complexes surrounding the rotor couple transmembrane $H^+$ or $Na^+$ flow to torque generation,[35, 36] and the resulting rotation is transmitted through the hook to the external helical filament to generate propulsion.[37] Its outstanding functionality stems from a complete working cascade: ion-motive energy input, sustained torque generation, cyclic rotary motion, and mechanical power transmission.[35, 36]

If this sequence of activities could be implemented using light, a nanoscale stator–rotor unit could be powered remotely. In this work, we realize this idea *in silico* with a nanoscale stator–rotor architecture in which the rotor position itself supplies the changing geometric polarity needed for sustained orbital motion (Figure 1a). Under circular polarization, the helicity-dependent force remains perpendicular to the instantaneous stator–rotor direction as the rotor moves, allowing the tangential force to perform nonzero work over a closed orbit. In contrast, linear polarization produces a conservative response with zero closed-cycle work. This coupling between optical helicity and structural polarity persists in both enclosed-shell and structurally eccentric stator configurations (Figure 1c), revealing the generality of this underlying physical principle. We further optimize this enclosed architecture for directed orbital motion with reduced optical heating while remaining robust to Brownian fluctuations (Figure 1d). Together, these results show that the stator–rotor geometry itself can supply the moving asymmetry needed to convert optical helicity into cyclic mechanical work.

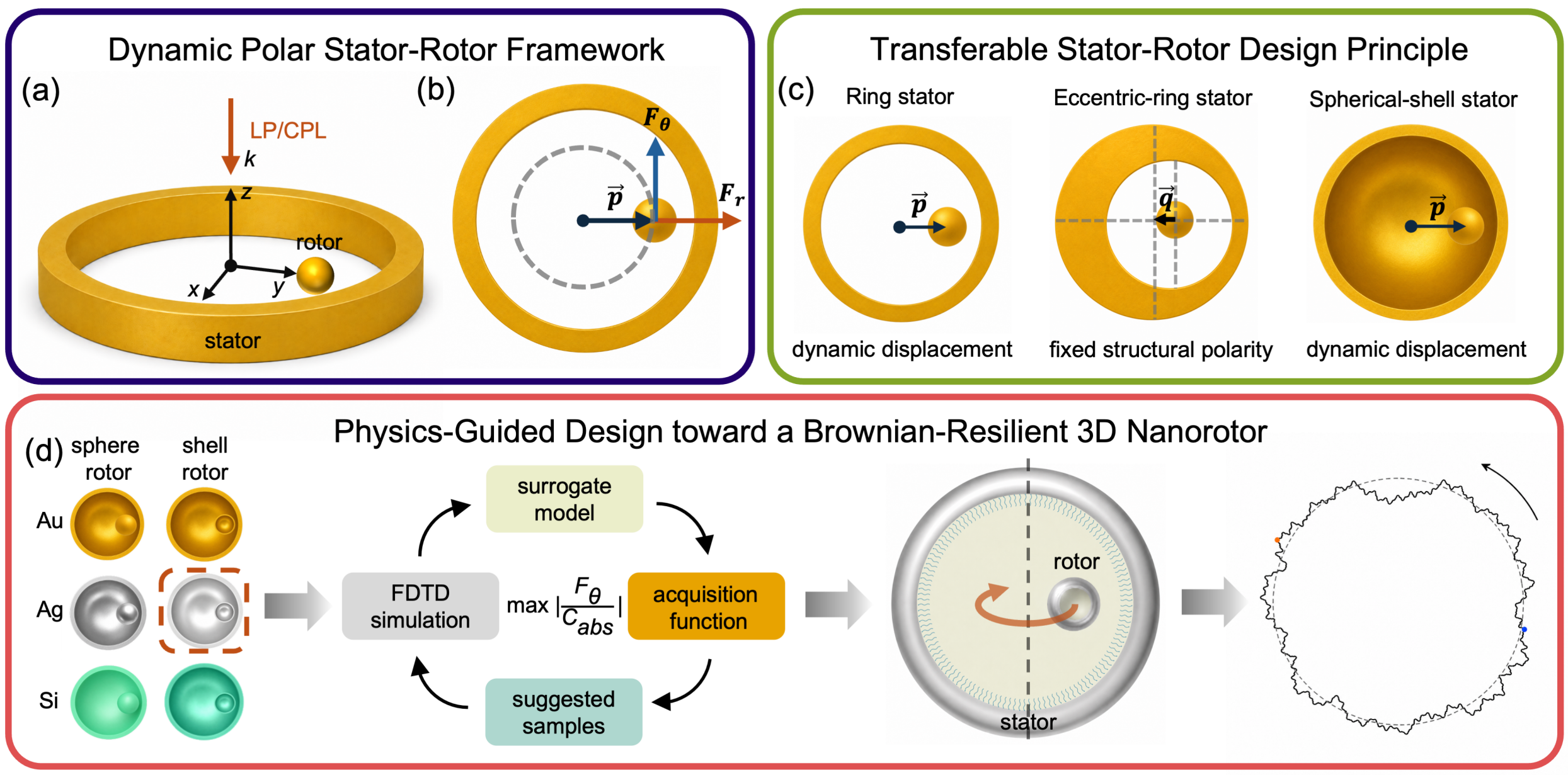

**Figure 1.** Conceptual overview of the optical stator–rotor framework and the device-design workflow. (a) Three-dimensional schematic of an achiral spherical rotor inside a ring stator under normally incident linearly polarized (LP) or circularly polarized (CPL) illumination. (b) Top-view definition of the dynamic stator-to-rotor displacement vector $\boldsymbol{p}$, the local radial–tangential directions defined by the instantaneous rotor displacement ($\boldsymbol{e}_r$ $\boldsymbol{e}_\theta$), and the corresponding force components $F_r$ *and* $F_\theta$. (c) Architectures used to test transferability of the helicity–polarity coupling: a ring and an enclosed spherical shell with a dynamic rotor displacement $\boldsymbol{p}$, and an eccentric ring with a fixed structural polarity $\boldsymbol{q}$. (d) Mechanism-guided design workflow comprising material and rotor-model screening, surrogate-assisted optimization, an enclosed rotor architecture, and Brownian-dynamics evaluation.

## Results and Discussion

### Off-Center Displacement Enables a Helicity-Dependent Tangential Force

To illustrate how rotor displacement changes the in-plane optical force response, we began with the simplest stator–rotor geometry: a spherical rotor inside a rotationally symmetric ring stator (Figure 1a,b). When the rotor is centered, all in-plane directions are equivalent. Moving the rotor off center breaks this symmetry: the line connecting the stator and rotor centers defines a polar axis $\boldsymbol{p}$. Because its direction is set by the rotor position, the polar axis would rotate with the rotor during orbital motion. The central question is therefore whether circularly polarized light can couple to this position-dependent asymmetry to generate a force perpendicular to the instantaneous stator–rotor direction.

We tested this idea using three-dimensional (3D) full-wave finite-difference time-domain (FDTD) simulations, with the optical force on the rotor obtained from the time-averaged Maxwell stress tensor (MST) (Supplementary Note S2).[38] The Au stator had an inner radius of 110 nm, an outer radius of 150 nm, and a height of 100 nm, while the Au spherical rotor had a radius of 40 nm. The *in silico* system was immersed in water and illuminated at normal incidence with an intensity of $1 \times 10^9$ W m$^{-2}$. For an off-center rotor, we define the instantaneous stator-to-rotor displacement vector $\boldsymbol{p}$, with $\rho = |\boldsymbol{p}|$. At each rotor position, the direction of $\boldsymbol{p}$ is the local radial direction, $\boldsymbol{e_r} = \boldsymbol{p}/\rho$, and rotating it counterclockwise by 90° gives the local tangential direction, $\boldsymbol{e_\theta} = \mathbf{J}\boldsymbol{e_r}$. The optical force is then resolved into the corresponding radial and tangential components $F_r$ and $F_\theta$.

We first tested whether breaking the center symmetry is required for a transverse force to appear. With the rotor at the ring center, the in-plane force vanishes across the spectrum under both LCP and RCP illumination (Figure 2b,d). Once the rotor is displaced by 30 nm along +*x*, however, a transverse *y*-directed force emerges and reaches approximately 30 fN near $\lambda^*$ = 589 nm (Figure 2b). For this geometry, the transverse direction coincides with the local tangential direction. At this displacement, the LCP and RCP tangential-force spectra are nearly identical in magnitude and line shapes but have opposite signs, whereas the corresponding radial-force spectra nearly overlap (Figure 2b,d). Thus, reversing optical helicity reverses $F_\theta$ while leaving $F_r$ essentially unchanged: $F_\theta$ is helicity-odd, whereas $F_r$ is helicity-even. Together, these results show that rotor displacement creates the in-plane structural polarity, and optical helicity then selects the sign of the tangential force.

Scanning the rotor along the $x$ axis from $d = -30$ to $+30$ nm at $\lambda^*$ = 589 nm reveals a nearly linear dependence of the transverse force on displacement (Figure 2c,e). The result shows that $F_y^{\mathrm{LCP}} = -0.943d$ fN and $F_y^{\mathrm{RCP}} = +0.943d$ fN, with $R^2$ = 0.9975 for both fits, where $d$ is expressed in nanometers. The longitudinal component shows the same near-linear dependence, with $F_x^{\mathrm{LCP}} = F_x^{\mathrm{RCP}} = 1.30d$ fN, with $R^2 = 0.9910$. Thus, moving the rotor from one side of the stator to the other flips both force components. At the same time, their magnitudes increase approximately linearly with $|d|$, and both vanish at the centered configuration.

The contrasting helicity dependences of the radial and tangential forces follow from the symmetry of the displaced geometry. The centered ring–sphere system has continuous in-plane rotational symmetry. Once the rotor is displaced, only one mirror plane remains: the plane containing $\boldsymbol{p}$ and the $z$ axis. Reflection across this plane leaves the radial direction unchanged but reverses the tangential direction. The radial and tangential force responses are therefore subject to different symmetry constraints. Therefore, any polarization contribution to $F_r$ must remain unchanged under this reflection, whereas a contribution to $F_\theta$ must change sign. At a fixed rotor displacement, Maxwell's equations are linear in the incident field amplitudes, while the time-averaged Maxwell stress tensor is quadratic in the resulting electromagnetic fields. Each in-plane force component can therefore be written as a Hermitian quadratic form of the local Jones vector $\boldsymbol{u} = (u_r, u_\theta)^{\mathrm{T}}$ (Supplementary Note S1):

$$F_i = \boldsymbol{u} \dagger \mathbf{K_i}\boldsymbol{u}, \qquad i \in \{r, \theta\}. \tag{1}$$

These symmetry requirements constrain the two response matrices to

$$\mathbf{K}_r = \begin{bmatrix} a & 0 \\ 0 & b \end{bmatrix}, \ \mathbf{K}_\theta = \begin{bmatrix} 0 & c \\ c^* & 0 \end{bmatrix} \tag{2}$$

where $a$ and $b$ are real and $c$ is generally complex. To connect these matrix forms to experimentally familiar polarization states, we rewrite the force responses in terms of the local Stokes parameters:

$$S_0 = |u_r|^2 + |u_\theta|^2, S_1 = |u_r|^2 - |u_\theta|^2,$$

$$S_2 = 2\mathrm{Re}(u_r * u_\theta), S_3 = 2\mathrm{Im}(u_r * u_\theta).$$

With these definitions, Equation 2 becomes

$$F_r = \alpha_0 S_0 + \alpha_1 S_1, \ F_\theta = \beta_2 S_2 + \beta_3 S_3, \tag{3}$$

where $\alpha_0 = (a+b)/2$, $\alpha_1 = (a-b)/2$, $\beta_2 = \mathrm{Re}(c)$, and $\beta_3 = -\mathrm{Im}(c)$ for the present helicity convention.

Equation 3 makes the different polarization dependences of the two force components explicit. The radial force contains a polarization-independent term $S_0$, together with a linear-polarization anisotropy term, $S_1$. The tangential force instead depends on $S_2$ and $S_3$, which describe the relative amplitude and phase between the radial and tangential field components. In particular, $S_3$ represents optical helicity.

For circular polarization, $\boldsymbol{u}_\sigma = (1, \mathrm{i}\sigma)^\mathrm{T}/\sqrt{2}$, where $\sigma = \pm 1$ denotes the two opposite optical helicities. In this case, $S_1 = S_2 = 0$ and $S_3 = \sigma$. Equation 3 therefore reduces to

$$F_r^{\mathrm{CPL}} = \alpha_0, \ \ F_\theta^{\mathrm{CPL}} = \beta_3 \sigma.$$

This result directly explains the helicity dependence observed in Figure 2b,d: reversing the optical helicity reverses $F_\theta$, while $F_r$ remains helicity independent.

The same symmetry argument also explains the displacement dependence in Figure 2c,e. At the centered configuration ($\rho = 0$), no in-plane direction is preferred, so the in-plane optical force must vanish. If the force varies smoothly away from the center, its leading term is therefore linear in displacement. This is consistent with the near-linear force trends observed in Figure 2c,e.

At a finite stator–rotor separation, the tangential force acts with a lever arm $\rho$ and therefore generates an orbital torque about the stator center,

$$\tau_z = (\boldsymbol{p} \times \boldsymbol{F})_z = \rho F_\theta.$$

At $d = 30$ nm and $\lambda^*$, the calculated orbital torque is $-894.3$ fN nm for LCP and $+894.3$ fN nm for RCP. This torque describes translation of the rotor around the stator axis, rather than spin about the rotor's own center.[5,13] However, a tangential force and the resulting orbital torque at a single rotor position do not by themselves guarantee sustained circulation. Equation 3 shows that linearly polarized light can also generate a tangential force through the $S_2$ term. The key question is therefore what the tangential force does over a complete orbit. We address this next by comparing the mechanical work produced under LP and CPL illumination.

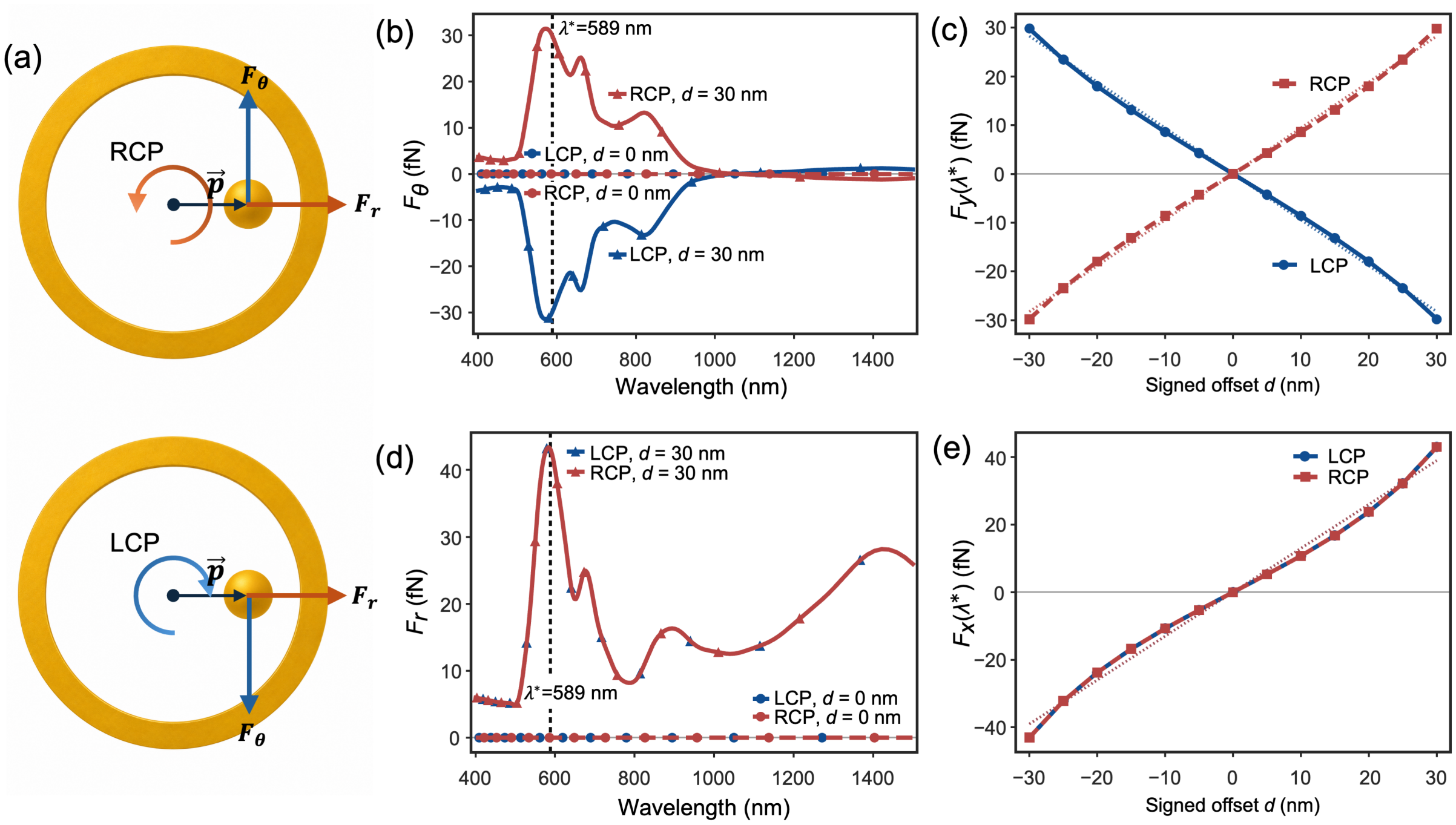

**Figure 2.** Helicity- and displacement-dependent optical forces in the ring–sphere system. (a) Schematic of the off-center ring–sphere configuration showing the local radial and tangential force directions under opposite circular-polarization handedness. (b) Transverse tangential force $F_\theta$ spectra under LCP and RCP illumination for centered ($d = 0$) and displaced ($d = 30\ nm$) rotor. The vertical marker indicates λ* = 589 nm, used for the displacement scans in (c, e). (c) Laboratory-frame transverse force $F_y$ as a function of signed rotor displacement $d$ at λ*. (d) Corresponding radial force $F_r$ spectra. (e) Laboratory-frame longitudinal force $F_x$ as a function of signed rotor displacement $d$ at λ*. Symbols in (c, e) denote FDTD calculations, and dotted lines denote linear fits.

## Circular Polarization Produces Nonzero Work over a Closed Orbit

To determine whether a tangential force is sufficient for sustained orbital driving, we first considered linearly polarized light as an informative control condition. Let $\theta$ denote the signed angle from the instantaneous radial direction $\boldsymbol{p}$ to the LP axis, measured positive in the local tangential direction (Figure 3a). In the local radial-tangential basis, the Jones vector is $\boldsymbol{u}_{\text{LP}} = (\cos\theta, \sin\theta)^{\text{T}}$, giving $S_0 = 1$, $S_1 = \cos(2\theta)$, $S_2 = \sin(2\theta)$, and $S_3 = 0$. Substituting these terms into Equation 3 gives

$$F_r^{\text{LP}} = \alpha_0 + \alpha_1\cos(2\theta), \qquad F_\theta^{\text{LP}} = \beta_2\sin(2\theta)$$

These expressions show that LP produces a twofold angular dependence in both force components. FDTD calculations confirm these predictions at $\lambda = 648.56$ nm and $\rho = 20$ nm: $F_\theta^{\text{LP}} = 20.0\sin(2\theta)$ fN ($R^2$ = 1.000), whereas $F_r^{\text{LP}} = 10.9 + 19.6\cos(2\theta)$ fN ($R^2$ = 1.000) (Figure 3b, c). Thus, LP can generate a substantial tangential force even without optical helicity.

The $\sin(2\theta)$ dependence directly explains why LP cannot drive continuous circular motion. As the rotor moves around a fixed-radius path, $F_\theta^{\text{LP}}$ changes sign every 90°, so the force drives the rotor in one direction over part of the orbit and in the opposite direction over the next. Because $F_\theta^{\text{LP}} \propto \sin(2\theta)$, the LP force can be written as the derivative of a periodic angular potential:

$$F_\theta = \frac{1}{\rho}\frac{dU}{d\theta}$$

Integrating the tangential force gives the angular potential $U(\theta) = -200\cos(2\theta)$ fN nm + constant under the conditions of Figure 3d. This periodic potential contains energy minima at specific angular positions, toward which the LP force tends to drive the rotor. The mechanical work supplied over one part of the orbit is canceled later in the same cycle, so the accumulated work returns to zero after one complete revolution (Figure 3d,f). LP can therefore select preferred angular positions, but it cannot continuously supply mechanical work around a closed orbit.

Circular polarization produces a fundamentally different force pattern. Unlike LP, CPL does not define a fixed in-plane polarization axis. Different azimuthal rotor positions are therefore equivalent under CPL. As a result, for a fixed helicity, $F_\theta$ retains the same sign as the rotor moves around the stator. Because that tangential direction itself turns with the rotor position, the driving force rotates with it and continues to push in the same circular motion direction throughout the orbit. The accumulated work therefore grows continuously with orbital angle (Figure 3e,f). For the prescribed counterclockwise path at $\lambda^* = 589\ nm$ and $\rho = 30\ nm$ in Figure 3, the calculated work per cycle is $W_{\text{cycle}}^{\text{LCP}} = -5619$ fN nm and $W_{\text{cycle}}^{\text{RCP}} = +5619$ fN nm. Reversing the optical helicity reverses the tangential force and, consequently, the preferred circular motion direction. Thus, the same stator–rotor geometry gives zero closed-cycle work under LP but sustained cyclic work under CPL.

Extending Equation 3 from a fixed rotor position to an arbitrary in-plane displacement $\boldsymbol{p}$ provides a common description of the LP and CPL responses. At the centered configuration, the in-plane force vanishes because no in-plane direction is preferred. For a smooth response near $\boldsymbol{p} = \boldsymbol{0}$, the leading displacement-dependent force is therefore linear in $\boldsymbol{p}$:

$$\boldsymbol{F}_\parallel = \boldsymbol{\Gamma}(\boldsymbol{u})\boldsymbol{p} + O(|\boldsymbol{p}|^2)\,.$$

Here, the form of matrix $\boldsymbol{\Gamma}$ is constrained by in-plane rotations: when the stator–rotor geometry is rotated, $\boldsymbol{p}$ and the resulting force must transform together. In two dimensions, $\boldsymbol{\Gamma}$ can be decomposed into three physically distinct parts. An isotropic part leaves $\boldsymbol{p}$ unchanged. An antisymmetric part rotates $\boldsymbol{p}$ by 90°. A symmetric anisotropic part is selected by linear polarization. These three contributions correspond respectively to total intensity, optical helicity, and the twofold LP axis. In the instantaneous local radial–tangential representation, the leading-order force law can be written as

$$\boldsymbol{F}_\parallel = [A(\lambda)\mathbf{I}_2 + B(\lambda)S_3\mathbf{J} + C(\lambda)\mathbf{Q}_\mathbf{L}]\boldsymbol{p} + O(|\boldsymbol{p}|^2) \tag{4}$$

where $\mathbf{I_2}$ is the two-dimensional identity matrix, $\mathbf{J}$ rotates an in-plane vector counterclockwise by 90°, and

$$\mathbf{Q}_\mathbf{L} = \begin{bmatrix} S_1 & S_2 \\ S_2 & -S_1 \end{bmatrix}$$

describes the axis selected by linear polarization. The coefficients $A$, $B$, and $C$ contain the wavelength-, material-, and geometry-dependent response amplitudes. Each term in Equation 4 has a direct mechanical meaning. The $A\mathbf{I}_2\boldsymbol{p}$ term points along $\boldsymbol{p}$ and gives the polarization-independent radial response. The $C\mathbf{Q}_\mathbf{L}\boldsymbol{p}$ term produces the twofold anisotropy imposed by linear polarization. Most importantly, the $BS_3\mathbf{J}\boldsymbol{p}$ term rotates the displacement direction by 90°: it is tangential at every rotor position, and its sign reverses with helicity. Resolving Equation 4 along the local radial and tangential directions gives

$$F_r = \rho[A(\lambda) + C(\lambda)S_1], \qquad F_\theta = \rho[C(\lambda)S_2 + B(\lambda)S_3].$$

These expressions are consistent with the local force relation in Equation 3 and connect directly to the angular dependence observed in Figure 3. For LP, the same coefficient $C$ produces the $\cos(2\theta)$ modulation in $F_r$ and the $\sin(2\theta)$ modulation in $F_\theta$, consistent with their nearly equal fitted amplitudes.

Equation 4 also explains why CPL can continuously supply mechanical work around a closed orbit. For CPL, $S_1 = S_2 = 0$, giving $F_r = A\rho$ and $F_\theta = BS_3\rho$. Because these force components are independent of the rotor azimuth, the configuration-space curl in polar coordinates is

$$\left(\nabla_{\boldsymbol{p}} \times \boldsymbol{F}_{\text{CPL}}\right)_z = \frac{1}{\rho}\frac{\partial(\rho F_\theta)}{\partial\rho} = 2B(\lambda)S_3. \tag{5}$$

A nonzero curl means that the force cannot be derived from a single-valued potential. The helicity-dependent CPL response is therefore nonconservative whenever $B \neq 0$.

For a circular orbit of radius $\rho$,

$$F_\theta = BS_3\rho, \qquad \tau_z = BS_3\rho^2,$$

and the work accumulated over one complete cycle is

$$W_{\text{cycle}}^{\text{CPL}} = \oint \boldsymbol{F} \cdot \mathrm{d}\boldsymbol{l} = 2\pi B(\lambda) S_3 \rho^2. \tag{6}$$

Equation 6 shows that the same helicity-dependent coefficient $B$ governs the tangential force, orbital torque, and closed-cycle work. Reversing the optical helicity reverses all three simultaneously. For LP, by contrast, $S_3 = 0$, $S_1 = \cos(2\theta)$, and $S_2 = \sin(2\theta)$, giving

$$F_r = \rho[A + C\cos(2\theta)], \qquad F_\theta = \rho C \sin(2\theta).$$

Because the signed angle $\theta$ decreases with increasing rotor azimuth, the corresponding configuration-space curl can be written as

$$\left(\nabla_{\boldsymbol{p}} \times \boldsymbol{F}_{\text{LP}}\right)_z = \frac{1}{\rho}\left[\frac{\partial(\rho F_\theta)}{\partial \rho} + \frac{\partial F_r}{\partial \theta}\right] = 0.$$

The same force field derives from the potential

$$U_{LP} = -\frac{1}{2}\rho^2[A + C\cos(2\theta)]$$

The LP response is therefore conservative, and the accumulated work returns to zero over any closed path. LP and CPL thus represent two distinct parts of the same stator–rotor force law: LP creates a periodic potential with preferred angular positions, whereas CPL provides a nonconservative tangential force that continuously supplies mechanical work around the orbit.

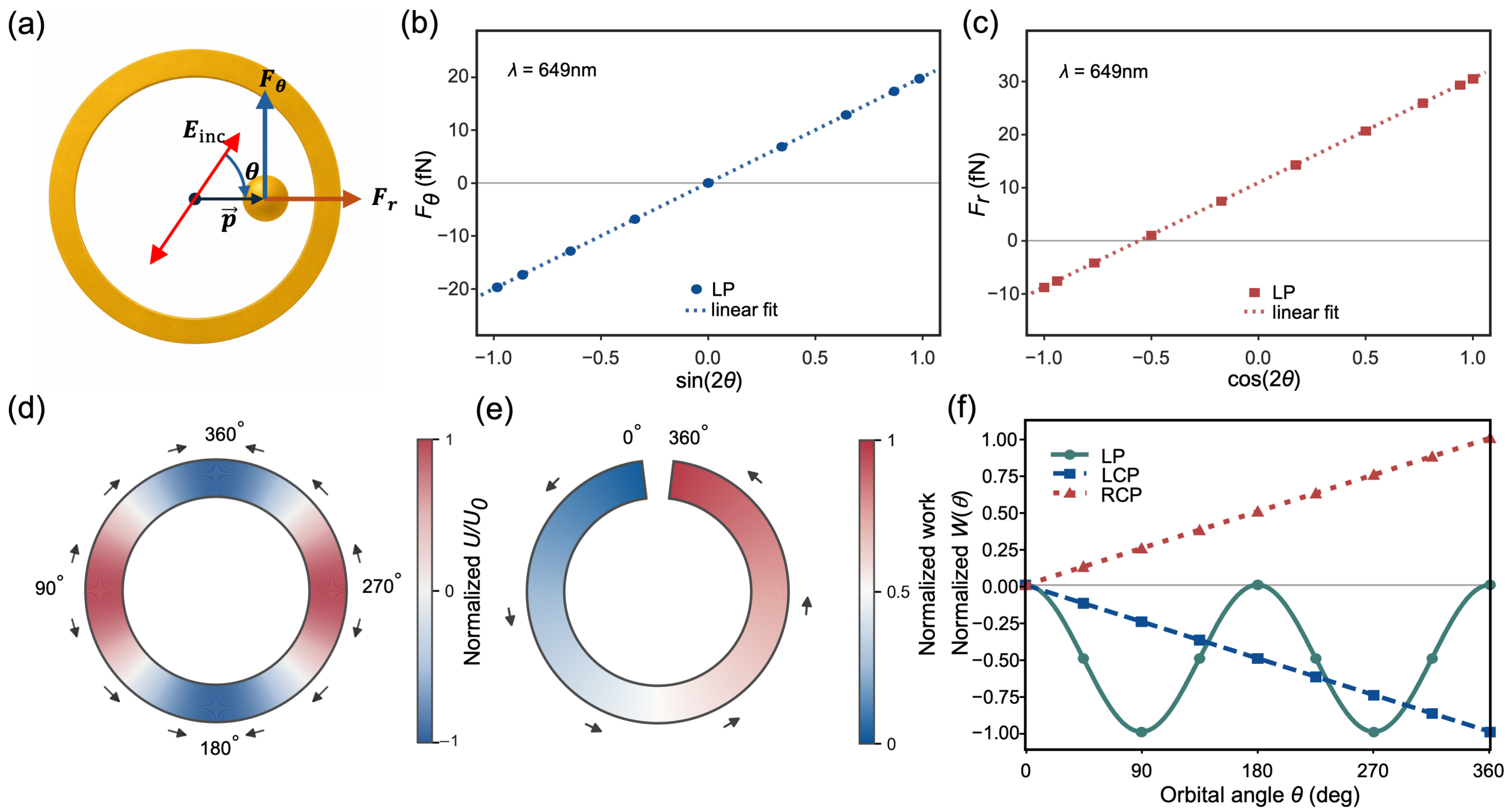

**Figure 3.** Polarization-dependent tangential forces and closed-cycle work in the ring–sphere system. (a) Schematic of the signed local polarization angle $\theta$, measured from the instantaneous radial direction $\boldsymbol{p}$ to the LP electric-field axis. (b) Tangential force as a function of $\sin(2\theta)$ at $\lambda = 649$ nm and $\rho = 20$ nm. (c) Radial force as a function of $\cos(2\theta)$ under the same conditions. Symbols in (b, c) denote FDTD calculations and dotted lines denote linear fits. (d) Normalized angular potential $(U(\theta)/U_0)$

obtained by integrating the LP tangential force along a fixed-radius orbit, where $U_0 = (U_{\max} - U_{\min})/2 = 200$ fN nm is the angular potential amplitude. (e) Accumulated work under LCP or RCP illumination along an unwrapped orbital coordinate from $0^\circ$ *to* $360^\circ$; the discontinuity between $0^\circ$ and $360^\circ$ represents the nonzero work accumulated upon returning to the same spatial position after one complete revolution. (f) Normalized accumulated work for LP, LCP, and RCP illumination over one orbital cycle. For CPL, the accumulated work is normalized by the magnitude of the total work per cycle, $|W_{\mathrm{cycle}}|$; for LP, it is normalized by the maximum absolute accumulated work within one cycle, $\max_\theta |W_{\mathrm{LP}}(\theta)|$.

## Helicity-Polarity Coupling Persists across Stator-Rotor Architectures

Having established the force mechanism in the minimal ring-sphere system, we next tested whether it persists in different stator geometries. The open ring is useful for isolating the underlying physics, but it does not physically enclose the rotor, so the rotor can still escape. We therefore replaced the ring with a closed spherical shell that preserves the rotational symmetry of the centered configuration while providing three-dimensional geometric confinement (Figure 4a). The shell had inner and outer radii of 180 and 230 nm, respectively, and contained a spherical rotor of radius 30 nm. Despite the substantial change in geometry, the same helicity-dependent tangential force response persists. When the rotor is centered, the in-plane force remains negligible; once the rotor is displaced from the shell center, a helicity-dependent tangential-force resonance emerges near $\lambda^* = 539$ nm (Figure 4b).

At $\lambda^*= 539$ nm, the shell reproduces the same displacement dependence observed in the ring-sphere system. The transverse force varies linearly with signed rotor displacement, following $F_y^{\mathrm{LCP}} = -7.76 \times 10^{-2} d$ fN and $F_y^{\mathrm{RCP}} = +7.76 \times 10^{-2} d$ fN, with $R^2 = 1.000$ for both fits (Figure 4c; $d$ in nanometers). Moving the rotor from one side of the stator to the other therefore reverses the transverse-force direction, just as in the ring geometry. In short, changing the stator from a ring to a shell mainly shifts the resonance wavelength and changes the force magnitude, while preserving this displacement-dependent response.

We then placed the rotor at different azimuthal positions along a complete orbit at fixed $\rho$. For a given helicity, the local tangential force remains nearly constant in magnitude and retains the same sign throughout the full $360^\circ$, while reversing sign when the optical helicity is reversed (Supplementary Note S4.1 and Figure S5). Thus, the helicity-dependent tangential force direction is perpendicular to the instantaneous stator-to-rotor polarity $\boldsymbol{p}$ at every tested rotor position. As the rotor moves around the stator, the force direction therefore rotates with it.

The ring and shell results show that rotor displacement can create the in-plane polarity required for the helicity-dependent tangential force. We next asked whether the displacement itself is essential, or whether any in-plane structural polarity is sufficient. To separate these two possibilities, we introduced the asymmetry directly into the stator. In the eccentric-ring geometry, the inner cavity is displaced relative to the outer stator while the rotor remains at the center of the local cavity (Figure 4d). This cavity offset defines a fixed structural polarity $\boldsymbol{q}$, even though the rotor itself is locally centered.

This built-in polarity produces a strong helicity-dependent transverse-force resonance near $\lambda^* = 852$ nm (Figure 4e). Reversing the structural offset reverses the force, with $F_\perp^{\mathrm{LCP}} = -1.84 d_q$fN and $F_\perp^{\mathrm{RCP}} = +1.84 d_q$ fN ($R^2 = 0.999$) (Figure 4f). Thus, rotor displacement is not required *per se* for the helicity-dependent response; the essential structural ingredient is an in-plane polarity. In the eccentric stator, the polarity $\boldsymbol{q}$ is built permanently into the structure, whereas in the ring and shell, it is supplied dynamically by the rotor displacement $\boldsymbol{p}$. The latter provides the orbital realization of this more general polarity-controlled response.

Having established that the force principle persists across different stator architectures, we next examined whether the shell implementation remains robust to geometric variations. Experimentally synthesized nanoshells often contain openings or incomplete closure.[39, 40] We therefore introduced an aperture into the shell as a representative structural imperfection. The main tangential-force resonance remains present in this opened-shell geometry over the tested rotor displacements (Figure 4g), and it persists as the spherical rotor radius is varied from 20 to 50 nm (Figure 4h).

The shell geometry provides a direct way to tune the optical response. Varying the shell inner radius from 150 to 180 nm and its thickness from 30 to 50 nm shifts the dominant force-peak wavelength from approximately 521 to 537.5 nm (Figure 4i; Supplementary Note S4.2–S4.4 and Figures S6–S8). In other words, the shell combines geometric robustness with spectral tunability, making it a suitable platform for the nanorotor design optimization described in the following.

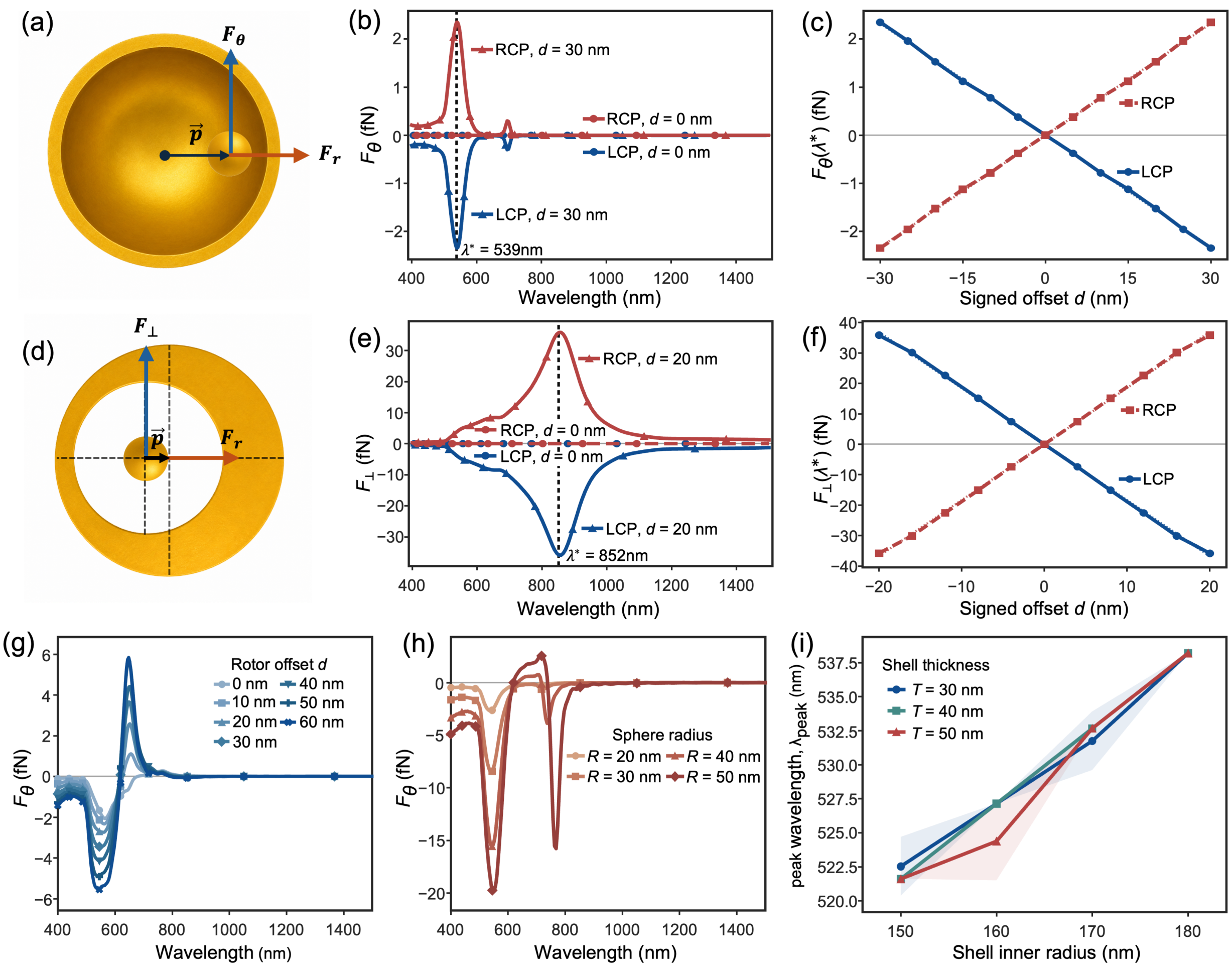

**Figure 4.** Robustness of helicity-controlled optical forcing across stator–rotor architectures and geometric variations. (a) Enclosing spherical-shell stator with a dynamically displaced spherical rotor; the shell has inner and outer radii of 180 and 230 nm, respectively, and the rotor radius is 30 nm. (b) Tangential force spectra under LCP and RCP illumination for centered ($d = 0$) and displaced ($d = 30$ nm) shell–rotor configurations. The vertical marker indicates $\lambda^* = 539$ nm, used for the displacement scan in (c). (c) Tangential force as a function of signed rotor displacement (d) at $\lambda^*$; symbols denote FDTD calculations and dotted lines denote linear fits. (d) Eccentric-ring stator with the rotor at the local cavity center; displacement of the inner cavity relative to the outer stator defines the fixed structural-polarity vector $\boldsymbol{q}$. (e) Tangential force spectra under LCP and RCP illumination for concentric and eccentric stator configurations. The vertical marker indicates $\lambda^* = 852$ nm, used for the structural-offset scan in (f). (f) Tangential force as a function of signed structural offset at $\lambda^*$; symbols denote FDTD calculations and dotted lines denote linear fits. (g) Tangential-force spectra for an opened-shell geometry, where a circular aperture (diameter 60 nm) is introduced in the shell with its center located in the rotor motion plane, as the rotor offset is varied from 0 to 60 nm. (h) Tangential-force spectra for spherical rotor radii from 20 to 50 nm. (i) Extracted force-peak wavelength as a function of shell inner radius (150–180 nm) and shell thickness (30–50 nm).

## Balancing Propulsion, Confinement, and Heating Enables Brownian Orbital Motion

A practical nanorotor design must combine sufficient tangential propulsion and stable radial confinement with moderate optical heating. For the device-design study, we adopted an enclosing stator with inner and outer radii of 180 and 210 nm, respectively, corresponding to the 30 nm-thick shell examined above. In the proposed design, poly(ethylene glycol) (PEG) chains grafted to the inner surface of the enclosing stator provide short-range steric repulsion as the rotor approaches the stator wall.[41, 42] This resulting inward repulsion counteracts the outward radial optical force, enabling a finite-radius force balance, while the helicity-dependent tangential force drives circular motion.

For the present device model, we represent the PEG-mediated interaction by an effective exponential soft-wall force,

$$F_r^{\mathrm{PEG}}(r) = -F_0 e^{\left(\frac{r-R_w}{\lambda_{\mathrm{PEG}}}\right)},$$

where $r$ is the radial displacement of the rotor center and $R_w = 60$ nm denotes the effective soft-wall position. $F_0$ sets the repulsion magnitude at $r = R_w$, whereas $\lambda_{\mathrm{PEG}}$ sets its characteristic decay length. The net radial force is then $F_r^{\mathrm{net}}(r) = F_r^{\mathrm{optical}}(r) + F_r^{\mathrm{PEG}}(r)$, with radial force balance defined by $F_r^{\mathrm{net}}(r) = 0$ (Supplementary Note S6.2).

With this radial force-balance model in place, we next optimized the nanorotor through material and rotor-architecture screening, geometric optimization, and subsequent photothermal and Brownian-dynamics validation.

Au-, Ag-, and Si-based stator–rotor material systems were first screened, with the stator and the optically active rotor material varied together. For each material system, both solid-sphere and nanoshell rotor architectures were evaluated; for the nanoshell rotors, the shell material was matched to that of the stator, while the $SiO_2$ core was kept fixed. The force-to-absorption ratio was evaluated at eligible tangential-force peaks, $|F_\theta(\lambda_{\mathrm{peak}})|/C_{\mathrm{abs}}(\lambda_{\mathrm{peak}})$. Only operating wavelengths with $F_r > 0$ were retained, consistent with the outward optical force required to balance the inward PEG-mediated repulsion. Among the tested configurations, the Ag nanoshell gave the highest score (1.554), followed closely by the Si sphere (1.542) and Ag sphere (1.477) (Figure 5a; Supplementary Note S5.1 and Figures S9 and S10). The Ag-nanoshell rotor consists of a $SiO_2$ core coated with an Ag shell, allowing the core radius and metal-shell thickness to be varied independently to tune its optical response.[43, 44] This geometric flexibility, together with its high force-to-absorption ratio, motivated its selection for further optimization.

Bayesian optimization (BO)[45] was then used to efficiently explore its geometric design space. For each geometry, $\lambda_{\mathrm{peak}}$ was defined as the eligible tangential-force maximum between 450 and 650 nm, subject to $F_r > 0$, and the corresponding ratio $|F_\theta(\lambda_{\mathrm{peak}})|/C_{\mathrm{abs}}(\lambda_{\mathrm{peak}})$ , served as the optimization objective. The best observed objective increased rapidly and approached a plateau after approximately 110 completed samples (Figure 5b), indicating that the search had efficiently identified the high-performing region of the design space.

We next considered tangential propulsion together with the outward radial response required for confinement. Because these two responses generally peak at different wavelengths, they must be compared at the same physical operating condition rather than by combining their separate spectral maxima. We therefore define each operating point as one geometry evaluated at one wavelength.

For each eligible operating point, we evaluated two absorption-normalized force metrics,

$$\eta_\theta = \frac{|F_\theta|}{C_{\mathrm{abs}}}, \qquad \eta_r = \frac{F_r}{C_{\mathrm{abs}}}, \qquad F_r > 0.$$

Here, $\eta_\theta$ measures tangential propulsion per unit absorption, whereas $\eta_r$ measures the outward radial response available to balance the confining wall interaction.

Because improving propulsion and confinement can compete, we compared the operating points using Pareto dominance.[46] An operating point is Pareto-nondominated when neither metric can be improved without reducing the other. Applying this criterion to the completed BO library produced an empirical Pareto front containing 20 operating points (Figure 5c; Supplementary Note S5.3). The original BO optimum lies near the high $\eta_\theta$ end of this front, as expected from the tangential-force objective used during the exploration stage.

To select a balanced operating point from this front, we normalized $\eta_\theta$ and $\eta_r$ and maximized the smaller of the two normalized scores. This maximin criterion selected Candidate 2 (C2; $R_{\text{core}} = 33.3\ \text{nm}, t_{\text{Ag}} = 25.2\ \text{nm}, R_{\text{out}} = 58.5\ \text{nm}$) at 533.8 nm. At this wavelength, C2 retains 96.45% of its maximum tangential efficiency and 90.63% of its maximum radial efficiency. We therefore selected C2 as the balanced representative for the subsequent validation.

The Pareto metrics are evaluated at the single displacement used for screening, so they do not yet establish stable radial confinement as the rotor moves. We therefore recalculated displacement-resolved radial and tangential force profiles for five representative Pareto candidates at their respective operating wavelengths (Supplementary Note S6 and Figure S14). For a common soft-wall model with $R_w = 60\ nm$, $F_0 = 0.40$ pN, and $\lambda_{\text{PEG}} = 0.50$ nm, all five candidates exhibited stable radial force balance near the effective soft-wall position (Figure S15). Their local radial stiffnesses differed substantially: the radial-endpoint candidate C1 reached 425 fN $\text{nm}^{-1}$, the balanced candidate C2 retained 374 fN $\text{nm}^{-1}$, and the original BO optimum C3 provided only 166 fN $\text{nm}^{-1}$(Table S2). The stiffness comparison confirms that strong tangential actuation does not necessarily provide strong radial restoring stiffness.

With the mechanical response established, we next assessed photothermal heating. At 534 nm and $1 \times 10^9\ \text{W m}^{-2}$, C2 gives a predicted maximum steady-state temperature rise of 20.90 K, compared with 56.22 K for the initial Au reference at the same incident intensity (Figure 5d). Both temperature rises were calculated for the complete stator–rotor structures at their respective operating wavelengths.

Static force balance and reduced heating, however, do not establish sustained orbital motion under thermal fluctuations. We therefore simulated overdamped Brownian dynamics using the displacement-resolved optical-force profiles, with the diffusion coefficient constrained by the fluctuation–dissipation relation, $D = \mu k_B T$.[47] For each condition, 32 independent trajectories were generated from independent Brownian-noise realizations. Across the five representative candidates and the tested soft-wall parameter range, all trajectories remained within the calculated 0–75 nm force domain and exhibited net circulation in the helicity-selected direction (Supplementary Note S8 and Figure S17). Among these candidates, C2 gives the narrowest central-90% radial distribution while retaining the second-highest mean circulation rate, further supporting it as the balanced design.

Having retained C2 as the optical design, we then optimized the two soft-wall parameters within this fixed geometry. Applying the same normalized maximin criterion to radial localization and circulation selected $\lambda_{\text{PEG}} = 0.55$ nm and $F_0 = 0.48$ pN, which were used for the final 32-trajectory ensemble.

With this parameter set, the steady-state occupancy from all 32 trajectories is concentrated mainly in an outer annulus near the effective soft wall, with Brownian fluctuation broadening the radial distribution (Figure 5e). The ensemble-averaged unwrapped orbital angle increases nearly linearly with time, corresponding to a mean angular speed of $1.350 \times 10^3\ \text{rad s}^{-1}$ (95% bootstrap confidence interval,[48] $1.302 \times 10^3$ to $1.395 \times 10^3\ \text{rad s}^{-1}$). Importantly, all 32 trajectories preserve the helicity-selected circulation direction (Figure 5f).

More broadly, this work establishes a transferable, simulation-guided workflow for nanorotor design, from material screening and multi-objective refinement to confinement, thermal, and Brownian-dynamics validation. This framework can be extended to other stator–rotor geometries and material systems.

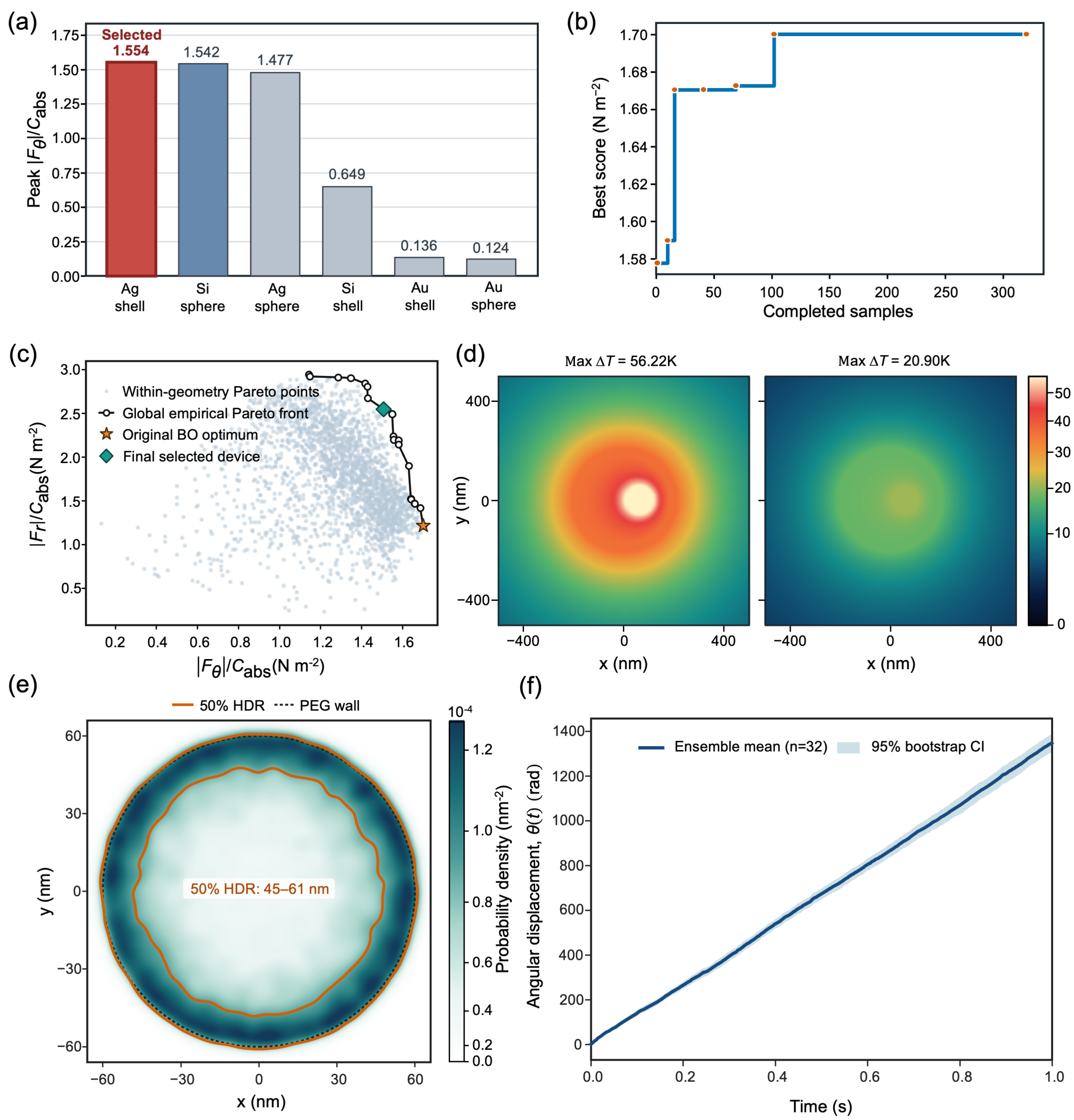

**Figure 5**. **Multi-objective nanorotor design and orbital dynamics under Brownian fluctuations.** (a) Screening of Au-, Ag-, and Si-based stator–rotor material systems with solid-sphere and nanoshell rotor architectures using the eligible peak tangential-force-to-absorption ratio, $|F_\theta|/C_{\mathrm{abs}}$. (b) Best observed objective as a function of completed samples during the single-objective Bayesian optimization of the Ag-nanoshell geometry. (c) Post-BO geometry–wavelength Pareto analysis of tangential-force efficiency, $|F_\theta|/C_{\mathrm{abs}}$, and outward radial-force efficiency, $F_r/C_{\mathrm{abs}}$ *with* $F_r > 0$. Light points denote within-geometry nondominated operating points, and open circles connected by the black line denote the empirical Pareto front among the 320 FDTD-evaluated geometries. The original BO optimum and balanced C2 operating point are highlighted. (d) Predicted steady-state temperature-rise maps for the initial Au reference and C2 Ag-nanoshell rotor at an incident intensity of $1.0 \times 10^9\ \mathrm{W\,m^{-2}}$, evaluated at their respective operating wavelengths. (e) Two-dimensional steady-state occupancy density for C2 obtained from all 32 fluctuation-dissipation-consistent Brownian trajectories over $t = 0.1 - 1.0s$. The orange contour encloses the 50% highest-density region,[49] and the dashed circle marks the nominal soft-wall position, $R_w = 60$ nm. (f) Ensemble-averaged unwrapped orbital angle for the same 32 trajectories; shading denotes the trajectory-level bootstrap 95% confidence interval.

## Conclusions

In summary, we theoretically identify a stator–rotor force principle in which a position-dependent geometric asymmetry couples optical helicity to sustained cyclic mechanical work. Displacing a spherical rotor from the stator center creates an in-plane polar axis that turns with the rotor position. Circular polarization couples to this moving polarity to generate a tangential force perpendicular to the instantaneous stator–rotor direction. As the rotor moves, this force remains tangential and continuously changes direction with the rotor position. The symmetry analysis reveals the origin of this behavior: the helicity-dependent response is antisymmetric and nonconservative, whereas the linear-polarization response is symmetric and conservative. This distinction explains why a tangential force alone does not guarantee sustained orbital driving—LP redistributes the rotor within a periodic potential, while CPL supplies meaningful work over repeated cycles. The persistence of the same helicity–polarity coupling in an enclosing shell and a structurally eccentric stator further shows that the underlying principle is not tied to a particular ring geometry or to rotor displacement as the only source of polarity.

The enclosing architecture also shows that this mechanism can be carried from a minimal physical model toward a constrained nanoscale rotor. Balancing propulsion, confinement, and optical heating identifies Ag-nanoshell operating conditions that retain directional orbital motion under Brownian fluctuations while substantially reducing photothermal heating relative to the initial Au reference. More broadly, the present framework shows that the direction of an optical mechanical force can be readily determined locally by the instantaneous stator–rotor geometry, rather than prescribed by a spatially structured external field. We envision that the stator–rotor architecture could be eventually realized for nanoscale mechanical output. An opening in the stator could act as an output port through which the orbiting rotor couples to a rigid nanoscale arm, for example, a DNA-origami element or a helical filament. The cyclic motion generated inside the cavity could then be transmitted to an external load. Such coupling would extend the present stator–rotor unit from an optically driven orbital rotor toward a polarization-programmable nanoscale rotary transducer that converts optical helicity into directional mechanical work.

## Methods

**Electromagnetic Simulations and Optical-Force Calculation.** Three-dimensional full-wave electromagnetic simulations were performed using Ansys Lumerical FDTD Solutions for stator–rotor structures immersed in water and illuminated at normal incidence by a plane wave. Optical forces on the rotor were calculated from the simulated electromagnetic fields using the time-averaged Maxwell stress tensor and resolved into the radial and tangential components defined in the main text. Mesh and integration-surface convergence were verified before production calculations. Detailed electromagnetic settings, material models, polarization conventions, force-integration procedures, and convergence tests are provided in the Supporting Information (Supplementary Note S2 and Figure S1).

**Photothermal Simulations.** Steady-state photothermal calculations used the FDTD-derived volumetric absorption density as the heat source in a three-dimensional finite-volume heat-conduction model. The temperature field was solved using the thermal properties of the constituent materials and surrounding water. The heat-source formulation, numerical discretization, boundary conditions, material parameters, solver validation, and convergence tests are provided in the Supporting Information (Supplementary Note S7 and Figure S16).

**Bayesian Optimization.** The Ag-nanoshell rotor was optimized over the $SiO_2$-core radius and Ag-shell thickness using a force-to-absorption objective obtained from FDTD/MST force and absorption spectra. A Gaussian-process surrogate with a Matérn kernel[50] and expected-improvement acquisition was initialized by Sobol sampling[51] and iterated over a total of 320 successfully scored geometries. Full design bounds, geometric constraints, objective definition, and optimization settings are provided in Supplementary Note S5.2.

**Post-BO Operating-Point Pareto Analysis.** The 320 FDTD-evaluated geometries were subsequently analyzed at the geometry–wavelength operating-point level. Each eligible geometry–wavelength pair was characterized by the tangential-force efficiency $|F_\theta|/C_{\mathrm{abs}}$ and outward radial-force efficiency $F_r/C_{\mathrm{abs}}, F_r > 0$. Spectrally dominated operating points were first removed within each geometry, followed by nondominated sorting across the remaining operating points. A normalized maximin criterion was used to identify a balanced Pareto candidate without introducing an explicit weighted objective. Complete definitions and candidate-selection procedures are provided in the Supporting Information (Supplementary Note S5 and Figure S12-S13).

**Brownian-Dynamics Simulations.** Rotor dynamics were modeled using a two-dimensional overdamped Langevin equation driven by the displacement-dependent radial and tangential optical forces. Translational mobility was described by Stokes drag and the diffusion coefficient satisfied the fluctuation–dissipation relation $D = \mu k_B T$; no empirical suppression of thermal noise was introduced. Radial confinement was represented by a short-range PEG-like soft-wall interaction, and stochastic performance was evaluated from ensembles of independent trajectories. Full force profiles, confinement model, integration parameters, parameter screening, and ensemble-statistical procedures are provided in the Supporting Information (Supplementary Note S6-8 and Figure S14-S20).

## Acknowledgment

We gratefully acknowledge the financial support from the National Natural Science Foundation of China (92356310, 22575200, 22572170) and Foundation of Westlake University. We thank Dr. Zhenhai Fu and Dr. Pan Wang of Zhejiang University for helpful discussions and for their assistance in assessing the physical soundness of this work.

## Supporting Information Available

Complete electrodynamic symmetry derivation; electromagnetic simulation details and numerical force-validation tests; additional CPL- and LP-resolved optical-force spectra; full-orbit and geometric-robustness analyses; material-screening and Bayesian-optimization details; photothermal-solver validation; and Brownian-dynamics model and parameter-sensitivity analyses.

## Conflict of Interest

The authors declare no competing financial interest.

## Data Availability Statement

The data supporting the findings of this study are available at Zenodo under DOI: 10.5281/zenodo.22763908.

# Supporting Information

# for

# "Sustained Orbital Motion Driven by Circularly Polarized Light in Nanoscale Stator–Rotor Architectures"

Shiye Du [a, b, c, †], Juanshu Wu [a, b, c, †], Xin Chen [e], Mouhong Lin [d, *], Hongyu Chen [a, b, c, *]

[a] Department of Chemistry, School of Science, Westlake University, Hangzhou 310030, China

[b] Research Center for Industries of the Future, Westlake University, Hangzhou 310030, China

[c] Institute of Natural Sciences, Westlake Institute for Advanced Study, Hangzhou 310024, China

[d] New Cornerstone Science Laboratory, State Key Laboratory of Extreme Photonics and Instrumentation, College of Optical Science and Engineering, Zhejiang University, Hangzhou, China

[e] Suzhou Laboratory, Suzhou 215123, China

[†] These authors contributed equally to this work.

[*] All correspondence should be addressed to linmouhong@zju.edu.cn and chenhongyu@westlake.edu.cn

## Note S1. Complete Electrodynamic Symmetry Derivation

This note provides the detailed derivation underlying the symmetry-constrained in-plane force law used in the main text. The derivation proceeds from the quadratic polarization dependence of the time-averaged optical force, applies the local mirror symmetry of a displaced stator–rotor configuration, and then promotes the local selection rules to a vector response that is covariant under in-plane rotations.

### S1.1 Quadratic polarization response at fixed displacement

At a fixed rotor displacement, Maxwell's equations are linear in the incident field amplitudes, whereas the time-averaged Maxwell stress tensor is quadratic in the electromagnetic fields. Consequently, each in-plane force component can be written as a Hermitian quadratic form of the incident Jones vector $\boldsymbol{u} = (u_r, u_\theta)^{\mathrm{T}}$ expressed in the instantaneous radial–tangential basis:

$$F_i = \boldsymbol{u} \dagger \mathbf{K}_{\mathrm{i}}\boldsymbol{u}, i \in \{r, \theta\} \tag{S1}$$

where $\mathbf{K}_i$ is a 2 × 2 Hermitian response matrix that depends on wavelength, geometry, material response, and the instantaneous stator–rotor configuration.

### S1.2 Constraint from the local radial mirror plane

The displaced ring–sphere configuration is invariant under reflection through the plane containing the displacement vector $\boldsymbol{p}$ and the $z$ axis. In the local $(r, \theta)$ basis, this reflection leaves radial quantities unchanged and reverses tangential quantities. Defining $\mathbf{M} = \begin{bmatrix} 1 & 0 \\ 0 & -1 \end{bmatrix}$, the Jones vector transforms as $\boldsymbol{u} \rightarrow \mathbf{M}\boldsymbol{u}$, while $F_r$ is even and $F_\theta$ is odd. The response matrices therefore obey

$$\mathbf{K}_r = \mathbf{M}\mathbf{K}_r\mathbf{M}, \mathbf{K}_\theta = -\mathbf{M}\mathbf{K}_\theta\mathbf{M} \tag{S2}$$

Together with Hermiticity, these constraints give

$$\mathbf{K}_r = \begin{bmatrix} a & 0 \\ 0 & b \end{bmatrix},\ \mathbf{K}_\theta = \begin{bmatrix} 0 & c \\ c^* & 0 \end{bmatrix} \tag{S3}$$

where a and b are real and c is generally complex.

### S1.3 Stokes decomposition and polarization selection rules

The local Stokes parameters are defined as

$$S_0 = |u_r|^2 + |u_\theta|^2 \tag{S4a}$$

$$S_1 = |u_r|^2 - |u_\theta|^2 \tag{S4b}$$

$$S_2 = 2\mathrm{Re}(u_r * u_\theta) \tag{S4c}$$

$$S_3 = 2\mathrm{Im}(u_r * u_\theta) \tag{S4d}$$

Substituting Eq. S3 into Eq. S1 yields

$$F_r = \alpha_0 S_0 + \alpha_1 S_1 \tag{S5a}$$

$$F_\theta = \beta_2 S_2 + \beta_3 S_3 \tag{S5b}$$

with $\alpha_0 = (a+b)/2$, $\alpha_1 = (a-b)/2$, $\beta_2 = \mathrm{Re}(c)$, and $\beta_3 = -\mathrm{Im}(c)$ for the helicity convention used in the main text. Thus, the radial channel contains the polarization-independent and local linear-anisotropy responses, whereas the tangential channel can arise only from interference between the local radial and tangential polarization components.

For circular polarization, $\boldsymbol{u}_\sigma = (1, \mathrm{i}\sigma)^{\mathrm{T}}/\sqrt{2}$, one has $S_1 = S_2 = 0$ and $S_3 = \sigma$. Therefore, the tangential response is helicity odd, $F_\theta = \beta_3\sigma$. Under the polarization convention adopted in the FDTD simulations, $\sigma = +1$ corresponds to LCP and $\sigma = -1$ corresponds to RCP.

For linear polarization, let $\theta$ denote the signed angle from the instantaneous radial direction to the LP axis, measured positive in the local tangential direction. Then $S_1 = \cos(2\theta)$, $S_2 = \sin(2\theta)$, and $S_3 = 0$, yielding the corresponding twofold radial and tangential responses.

### S1.4 Leading-order spatial response near the centered configuration

The centered configuration $(\boldsymbol{p} = 0)$ has no in-plane polar axis and therefore no in-plane force. Assuming analyticity about the centered configuration, the leading spatial response is linear in the displacement vector:

$$\boldsymbol{F}_\parallel = \boldsymbol{\Gamma}(\boldsymbol{u})\boldsymbol{p} + O(|\boldsymbol{p}|^2) \tag{S6}$$

where $\boldsymbol{\Gamma}$ is a real $2 \times 2$ response matrix whose polarization dependence is quadratic in the incident field. Any real $2 \times 2$ matrix can be decomposed uniquely into an isotropic part, an antisymmetric part, and a symmetric-traceless part.

### S1.5 Rotational covariance and the leading-order force law

The polarization bilinears possess the same transformation sectors. $S_0$ is a scalar under in-plane rotations, $S_3$ is the helicity pseudoscalar, and the pair $(S_1, S_2)$ forms the linear-polarization symmetric-traceless sector. Introducing

$$\mathbf{J} = \begin{bmatrix} 0 & -1 \\ 1 & 0 \end{bmatrix} \tag{S7}$$

$$\mathbf{Q_L} = \begin{bmatrix} S_1 & S_2 \\ S_2 & -S_1 \end{bmatrix} \tag{S8}$$

In the instantaneous local radial–tangential representation, rotational covariance and mirror symmetry constrain the leading-order response, for normalized illumination $S_0 = 1$, to the form

$$\boldsymbol{F}_{\parallel} = [A(\lambda)\mathbf{I}_2 + B(\lambda)S_3\mathbf{J} + C(\lambda)\mathbf{Q_L}]\boldsymbol{p} + O(|\boldsymbol{p}|^2) \tag{S9}$$

where $A$, $B$, and $C$ contain the wavelength-, material-, and geometry-dependent response amplitudes. In this local representation, $\boldsymbol{p} = (\rho, 0)^{\mathrm{T}}$, so Eq. S9 gives

$$F_r = \rho[A + CS_1] \tag{S10a}$$

$$F_\theta = \rho[CS_2 + BS_3] \tag{S10b}$$

The same coefficient $C$ governs both the LP radial-anisotropy and tangential twofold channels.

**S1.6 Conservative and nonconservative sectors**

For circular polarization, $S_1 = S_2 = 0$, so Eq. S10 gives

$$\begin{aligned} F_r &= A\rho \\ F_\theta &= BS_3\rho \end{aligned} \tag{S11a}$$

Because these force components are independent of the rotor azimuth, the two-dimensional configuration-space curl in polar coordinates is

$$\begin{aligned} \left(\nabla_{\boldsymbol{p}} \times \boldsymbol{F}_{\mathrm{CPL}}\right)_z &= \frac{1}{\rho}\frac{\partial(\rho F_\theta)}{\partial\rho} \\ &= 2B(\lambda)S_3 \end{aligned} \tag{S11b}$$

A nonzero curl means that the force cannot be derived from a single-valued potential. The CPL response is therefore nonconservative whenever $B \neq 0$.

For a circular orbit of radius $\rho$, the associated orbital torque and work per cycle are

$$\tau_z = (\boldsymbol{p} \times \boldsymbol{F}_{\mathrm{CPL}})_z = BS_3\rho^2 \tag{S12}$$

$$W_{\mathrm{cycle}}^{\mathrm{CPL}} = \oint \boldsymbol{F} \cdot \mathrm{d}\boldsymbol{l} = 2\pi B(\lambda)S_3\rho^2 \tag{S13}$$

Thus, reversing optical helicity reverses the tangential force, orbital torque, and closed-cycle work simultaneously.

For linear polarization, $S_3 = 0$, $S_1 = \cos(2\theta)$, and $S_2 = \sin(2\theta)$, giving

$$\begin{aligned} F_r &= \rho[A + C\cos(2\theta)] \\ F_\theta &= \rho C\sin(2\theta) \end{aligned} \tag{S14a}$$

Let $\varphi$ denote the rotor azimuth measured in the positive tangential direction. For a fixed LP axis, $\mathrm{d}\theta = -\mathrm{d}\varphi$. Therefore, the standard polar-coordinate curl can be written as

$$
\begin{aligned}
&\left(\nabla_{\boldsymbol{p}} \times F_{\mathrm{LP}}\right)_z \\
&= \frac{1}{\rho}\left[\frac{\partial(\rho F_\theta)}{\partial \rho} - \frac{\partial F_r}{\partial \varphi}\right] \\
&= \frac{1}{\rho}\left[\frac{\partial(\rho F_\theta)}{\partial \rho} + \frac{\partial F_r}{\partial \theta}\right] \\
&= 0
\end{aligned}
\tag{S14b}
$$

The same force field derives from the potential

$$
U_{\mathrm{LP}}(\rho,\theta) = -\frac{1}{2}\rho^2[A + C\cos(2\theta)] \tag{S14c}
$$

Indeed,

$$
\begin{aligned}
F_r &= -\frac{\partial U_{\mathrm{LP}}}{\partial \rho}, \\
F_\theta &= -\frac{1}{\rho}\frac{\partial U_{\mathrm{LP}}}{\partial \varphi} = \frac{1}{\rho}\frac{\partial U_{\mathrm{LP}}}{\partial \theta},
\end{aligned}
\tag{S14d}
$$

consistent with the signed-angle convention above. The leading-order LP response is therefore conservative and performs zero work over any closed path.

Higher-order terms represented by $O(|\boldsymbol{p}|^2)$ are not included in this leading-order symmetry statement.

## Note S2. Electromagnetic Simulation and Optical-Force Calculation

### S2.1 Electromagnetic Simulation Setup

Three-dimensional full-wave electromagnetic simulations were performed using Ansys Lumerical FDTD Solutions. The stator–rotor structures were immersed in water and illuminated at normal incidence using a total-field/scattered-field (TFSF) source. Unless otherwise stated, the incident intensity was $1 \times 10^9$ W m$^{-2}$ and wavelength-resolved force and absorption spectra were evaluated at 200 uniformly spaced frequency points spanning the frequency interval corresponding to wavelengths from 400 to 1500 nm.

Linearly polarized illumination was implemented using the corresponding in-plane electric-field component. Circularly polarized illumination was generated by superposing two orthogonal in-plane linear-polarization components with equal amplitudes and a relative phase of $\pm\pi/2$. Equivalently, the incident circularly polarized field can be written as

$$\boldsymbol{E}_\sigma = \frac{E_0}{\sqrt{2}}(\hat{\boldsymbol{x}} + i\sigma\hat{\boldsymbol{y}})$$

where $\sigma = \pm 1$ denotes the two opposite optical helicities. Under the polarization convention used in the simulations, $\sigma = +1$ corresponds to LCP and $\sigma = -1$ to RCP.

For the ring–sphere calculations used to establish the optical-force mechanism in the main text, a mesh step of 1.3 nm was used for the production simulations. The adequacy of this discretization was verified by systematic mesh-convergence calculations, as described in Section S2.3. The simulation boundaries were terminated by perfectly matched layers (PMLs), with sufficient separation between the nanostructure and the boundaries to suppress boundary-induced artifacts. The wavelength-dependent optical properties of Au, Ag, Si, and $SiO_2$ were described using the dispersive material models implemented in Ansys Lumerical FDTD Solutions. Au and Ag were modeled using the Johnson–Christy optical-constant data,[1] whereas Si and $SiO_2$ (glass) were modeled using the Palik data.[2, 3] The complex refractive indices used by the FDTD solver were obtained from the corresponding fitted dispersive material models over the simulated spectral range.

### S2.2 Maxwell-Stress-Tensor Force Calculation

The time-averaged optical force acting on the rotor was evaluated by integrating the Maxwell stress tensor over a closed surface enclosing the rotor,[4]

$$\boldsymbol{F} = \oint_S \langle \mathbf{T} \rangle \cdot \hat{\boldsymbol{n}}, \mathrm{d}S$$

where $S$ denotes the closed integration surface and $\hat{\boldsymbol{n}}$ is its outward unit normal. In the homogeneous surrounding medium, the time-averaged Maxwell stress tensor was evaluated from the simulated electric and magnetic fields as

$$\langle \mathbf{T} \rangle = \frac{1}{2} \mathrm{Re} \left[ \varepsilon_b \boldsymbol{E}\boldsymbol{E}^* + \mu_b \boldsymbol{H}\boldsymbol{H}^* - \frac{1}{2} (\varepsilon_b |\boldsymbol{E}|^2 + \mu_b |\boldsymbol{H}|^2) \mathbf{I} \right]$$

where $\varepsilon_b$ and $\mu_b$ are the permittivity and permeability of the surrounding medium, respectively, and $\mathbf{I}$ is the identity tensor.

A rectangular integration box surrounding the rotor was used for the numerical surface integration. The entire integration surface was located within homogeneous water, enclosed only the rotor, and did not intersect the stator. The production integration surface was positioned 5 mesh cells away from the rotor surface. The calculated force was subsequently resolved into the instantaneous radial and tangential directions defined by the stator-to-rotor displacement vector $\boldsymbol{p}$,

$$F_r = \boldsymbol{F} \cdot \hat{\boldsymbol{e}}_r, \qquad F_\theta = \boldsymbol{F} \cdot \hat{\boldsymbol{e}}_\theta$$

For signed displacement scans along the x axis, the corresponding laboratory-frame components $F_x$ and $F_y$ were retained, with $F_x$ and $F_y$ representing the local radial and tangential components, respectively.

### S2.3 Numerical convergence and force-method validation

To verify that the optical forces reported for the ring–sphere system were not sensitive to the numerical discretization or to the implementation of the force calculation, we performed three complementary numerical tests: electromagnetic-mesh convergence, cross-validation between two independent force-evaluation methods, and convergence with respect to the Maxwell-stress-tensor (MST) integration surface.

#### S2.3.1 FDTD Mesh Convergence

Mesh convergence was evaluated using a representative off-center ring–sphere configuration in water under the same incident intensity used in the main calculations, $I = 1 \times 10^9$ W, m$^{-2}$. All geometrical parameters, source settings, material models, and force-integration settings were kept unchanged while the local electromagnetic mesh step was varied from 0.7 to 1.5 nm.

The principal force spectra and resonance positions remain stable over the tested mesh range (Figure S1a,b). At the operating wavelength used for the main-text analysis, the deviations of $F_x$ ($F_r$) and $F_y$($F_\theta$) obtained with the 1.3 nm mesh from the finest-mesh results are 2.73% and 2.32%, respectively. A mesh step of 1.3 nm was therefore used for the production ring–sphere calculations.

### S2.3.2 Cross-Validation of MST and Volumetric Force Calculations

The MST force was independently compared with a volumetric optical-force calculation using the same electromagnetic field solutions. The two methods reproduce the same spectral positions, force signs, and force magnitudes over the wavelength range relevant to the main-text analysis (Figure S1c,d). The MST formulation was therefore used for all production force calculations.

### S2.3.3 MST Integration-Surface Convergence

We additionally examined whether the calculated force depended on the location of the closed surface used for Maxwell-stress-tensor integration. The integration box was expanded outward from the rotor surface by 3–7 mesh cells while keeping the electromagnetic mesh and all physical parameters fixed. In every case, the integration surface remained entirely within the surrounding homogeneous medium, enclosed only the rotor, and did not intersect the stator.

As shown in Figure S1c,d, the calculated force spectra are essentially unchanged over the entire tested range of integration-surface offsets. Both $F_x$ and $F_y$, including their resonance positions and amplitudes, remain stable as the MST box is moved away from the particle surface. This confirms that the reported optical forces are insensitive to the precise placement of the MST integration surface within the tested range.

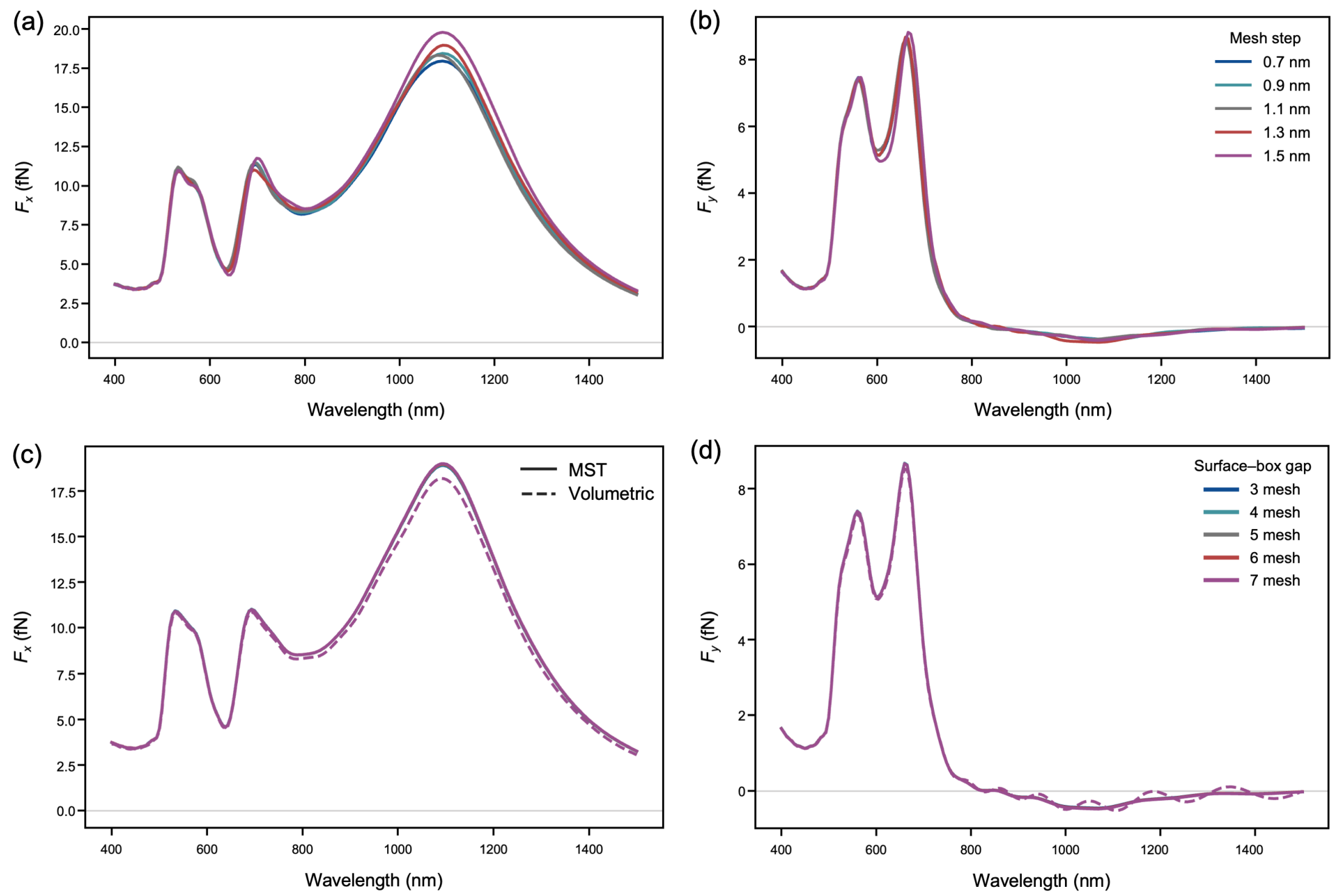

**Figure S1. Numerical validation of the optical-force calculations for the ring–sphere system.** (a,b) Mesh-convergence tests for the laboratory-frame force components $F_x$ and $F_y$ for mesh steps of 0.7-1.5 nm. (c,d) Cross-validation of Maxwell-stress-tensor and volumetric force calculations and dependence on the MST integration-surface offset. The production calculations used a 1.3 nm mesh and a five-cell MST-surface offset.

## Note S3. Additional Optical-Force Spectra of the Ring–Sphere System

### S3.1 Displacement-Resolved CPL Force Spectra

To complement the representative spectra and signed-displacement scans in Figure 2 of the main text, wavelength-resolved radial and tangential forces were calculated for the full set of signed rotor displacements under both LCP and RCP illumination (Figure S2).

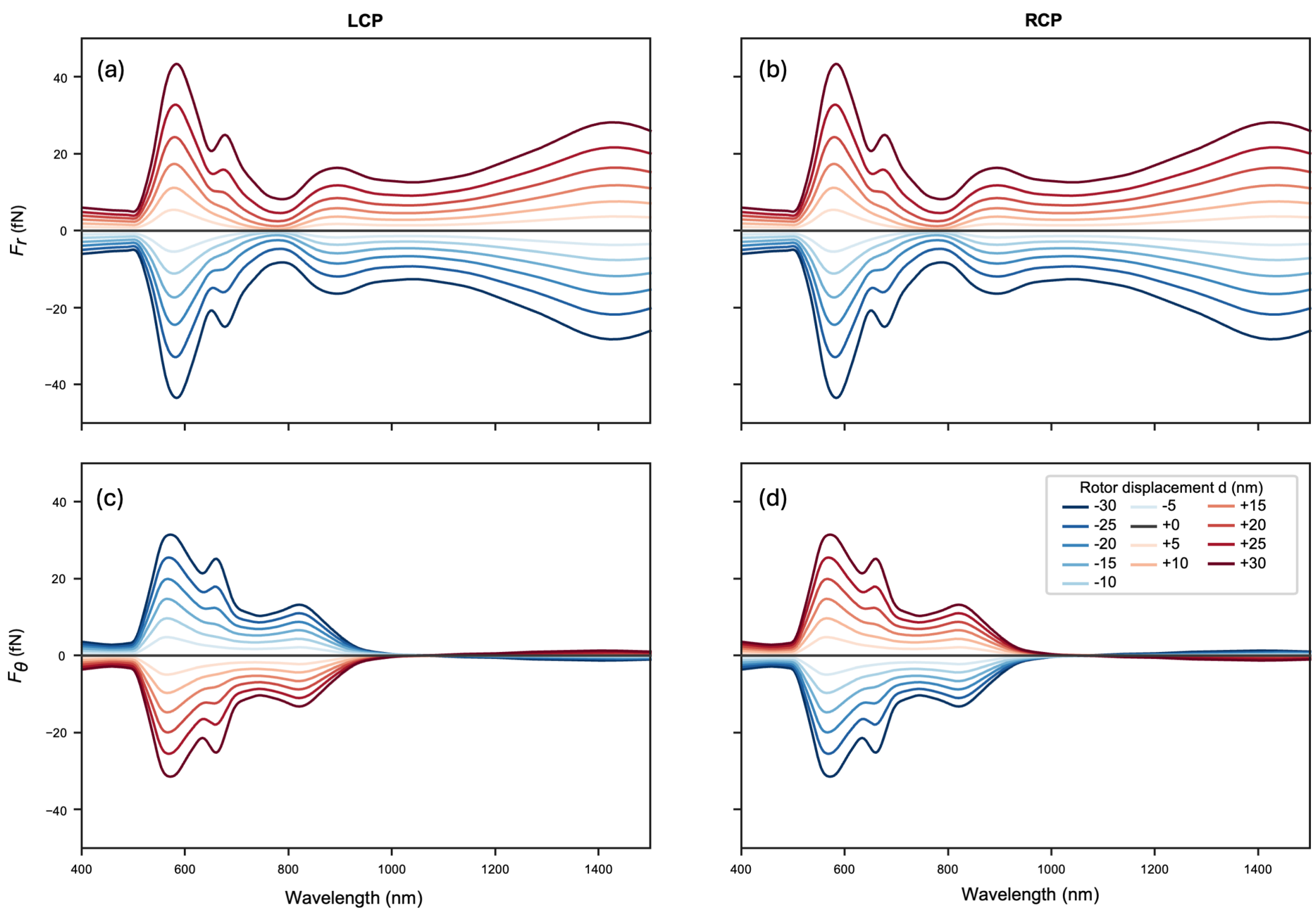

**Figure S2. Displacement-resolved optical-force spectra of the ring–sphere system under circularly polarized illumination.** (a,b) Radial-force spectra $F_r$ under (a) LCP and (b) RCP illumination. (c,d) Corresponding tangential-force spectra $F_\theta$. Curves correspond to signed rotor displacements $d = -30\ to + 30$ nm, with negative and positive displacements shown in blue and red, respectively. All other simulation parameters are identical to those used in Figure 2 of the main text.

### S3.2 Polarization-Angle-Resolved LP Force Spectra

The wavelength-resolved force spectra underlying the LP angular dependence in Figure 3 of the main text were calculated for polarization angles from $0^\circ$ to $170^\circ$ in $10^\circ$ increments. The corresponding tangential and radial force spectra are shown in Figures S3 and S4, respectively.

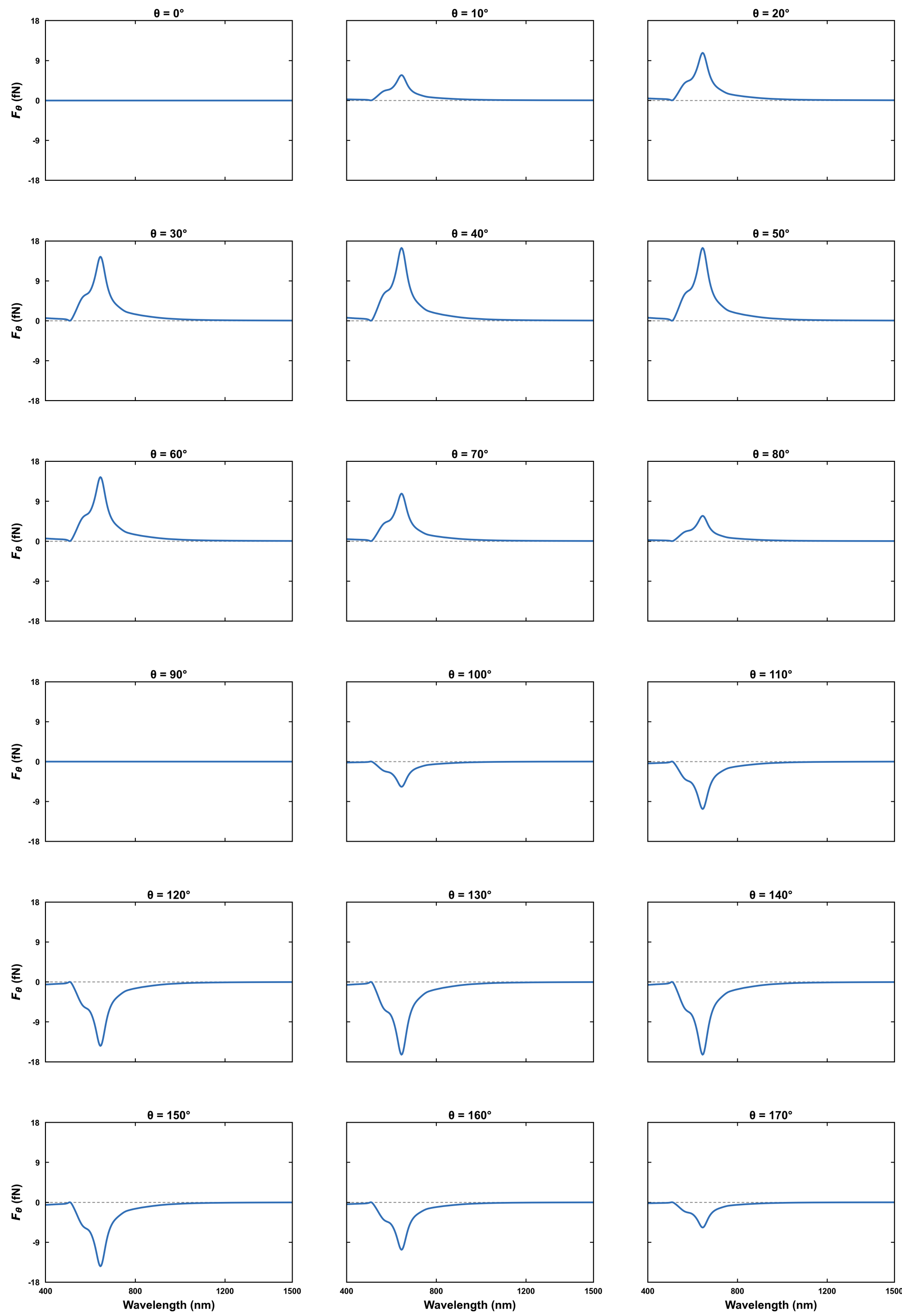

**Figure S3. Wavelength-resolved tangential-force response under linearly polarized illumination.** Tangential-force spectra $F_\theta$ calculated for polarization angles $\theta = 0° -$ $170°$ in $10°$ increments at a rotor displacement of $d = 20$ nm. **Here, $\theta$ is the signed angle from the local radial direction to the incident linear-polarization axis, measured positive in the local tangential direction.**

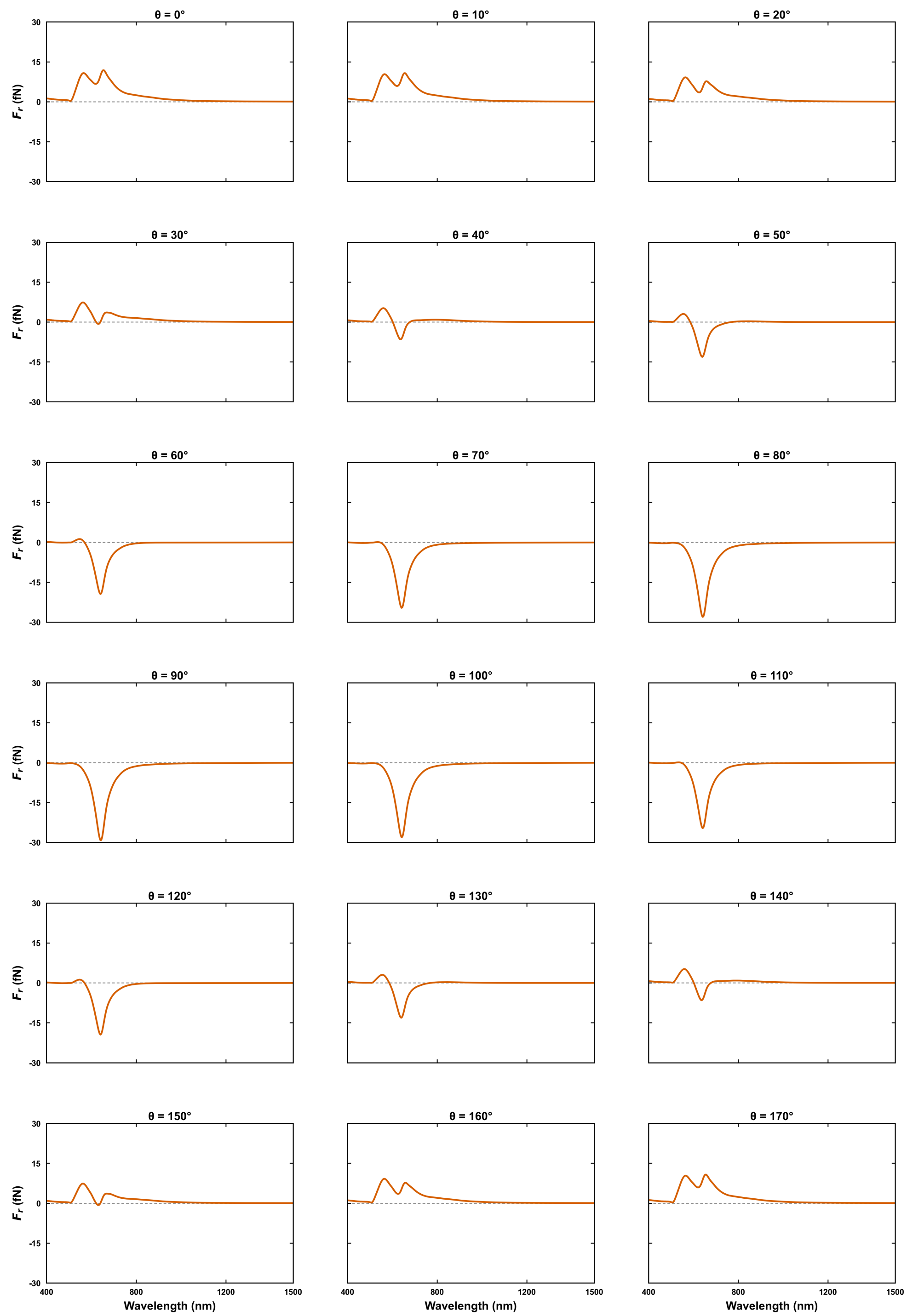

**Figure S4. Wavelength-resolved radial-force response under linearly polarized illumination.** Radial-force spectra $F_r$ for polarization angles $\theta = 0° - 170°$ in $10°$ increments at a rotor displacement of $d = 20$ nm. The angle $\theta$ is defined as in Figure S3.

## Note S4. Additional Validation across Stator–Rotor Architectures

### S4.1 Full-Orbit Azimuthal Validation in the Shell–Sphere Architecture

To verify that the tangential-force direction follows the instantaneous rotor position rather than a fixed laboratory axis, the displaced sphere was sampled over a complete orbit inside the enclosing shell at fixed radial displacement and wavelength (Figure S5). The small residual azimuthal variation is consistent with the finite Cartesian discretization of the numerical model.

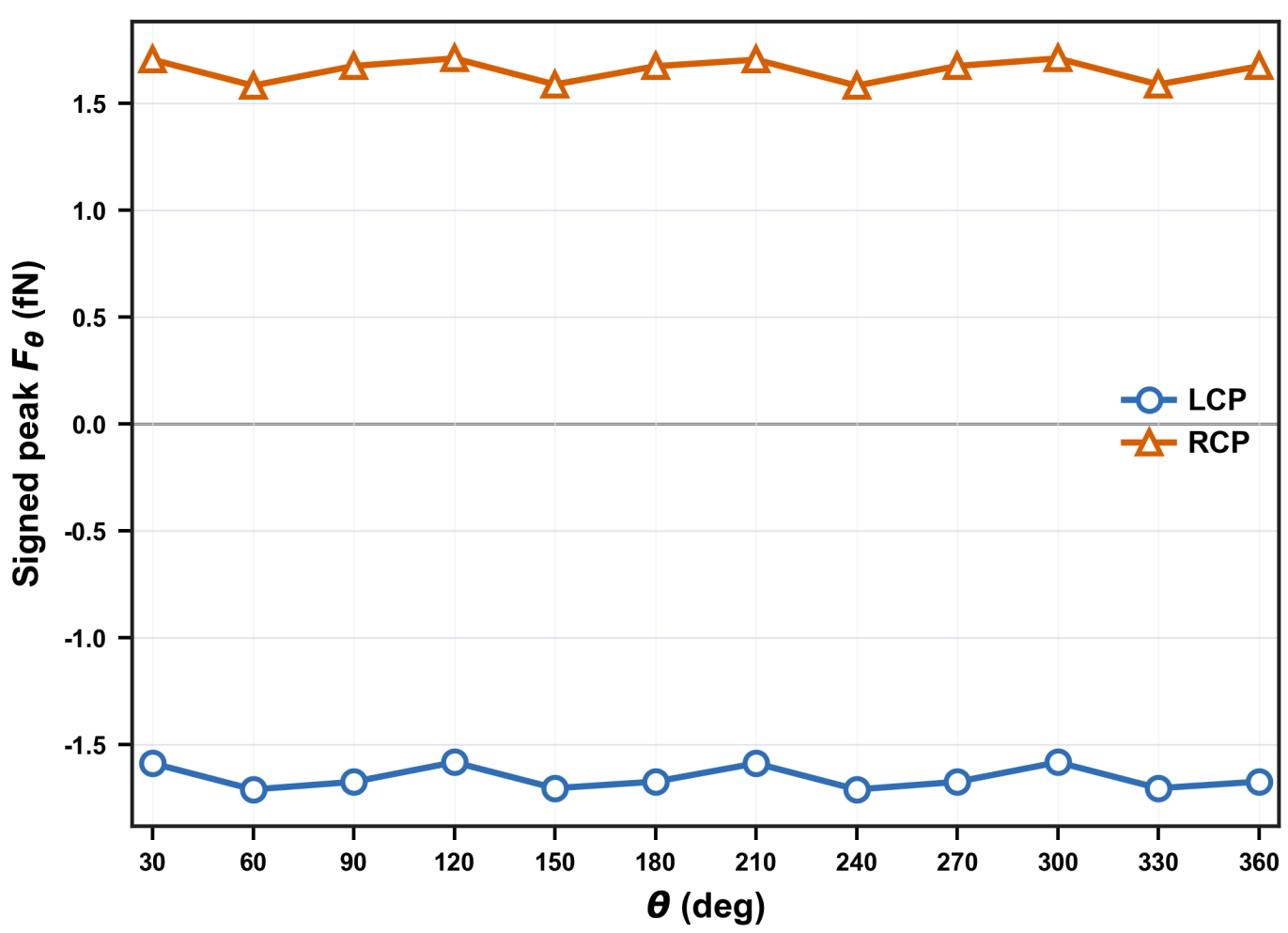

**Figure S5. Full-orbit azimuthal validation in the shell–sphere architecture.** Tangential force $F_\theta$ as a function of rotor azimuthal position $\phi$ over a complete orbit under LCP and RCP illumination at $\lambda = 523\ nm$. The geometry is identical to that in Figure 4a–c of the main text. $F_\theta$ remains nearly independent of azimuth and reverses sign with optical helicity.

### S4.2 Eccentric Ring–Sphere Response under Circular Polarization

For the eccentric-ring geometry in Figure 4d–f of the main text, the sphere was retained at the center of the inner cavity while the signed eccentricity $d$ was varied by displacing the inner cavity relative to the outer stator. Figure S6 shows the corresponding wavelength-resolved tangential-force spectra under LCP and RCP illumination for the tested eccentricities.

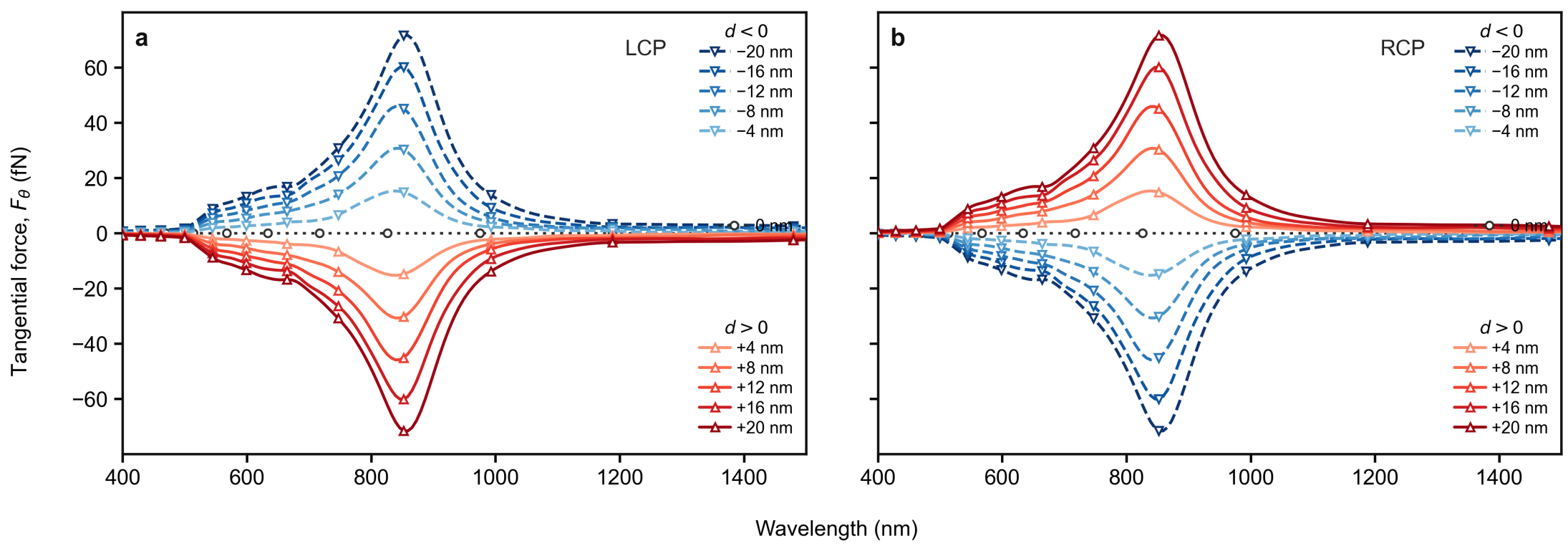

**Figure S6. Tangential-force spectra of the eccentric ring–sphere geometry under circularly polarized illumination.** Tangential-force spectra under (a) LCP and (b) RCP illumination for different signed stator eccentricities $d$. The rotor remains at the center of the inner cavity, while $d$ denotes the displacement of the inner-cavity center relative to the outer-stator center; $d = 0$ denotes the concentric stator.

### S4.3 Helicity Reversal in the Opened-Shell Architecture

To complement the opened-shell displacement scan in Figure 4g of the main text, wavelength-resolved tangential-force spectra were calculated under both LCP and RCP illumination for the same set of rotor displacements. The results are shown in Figure S7.

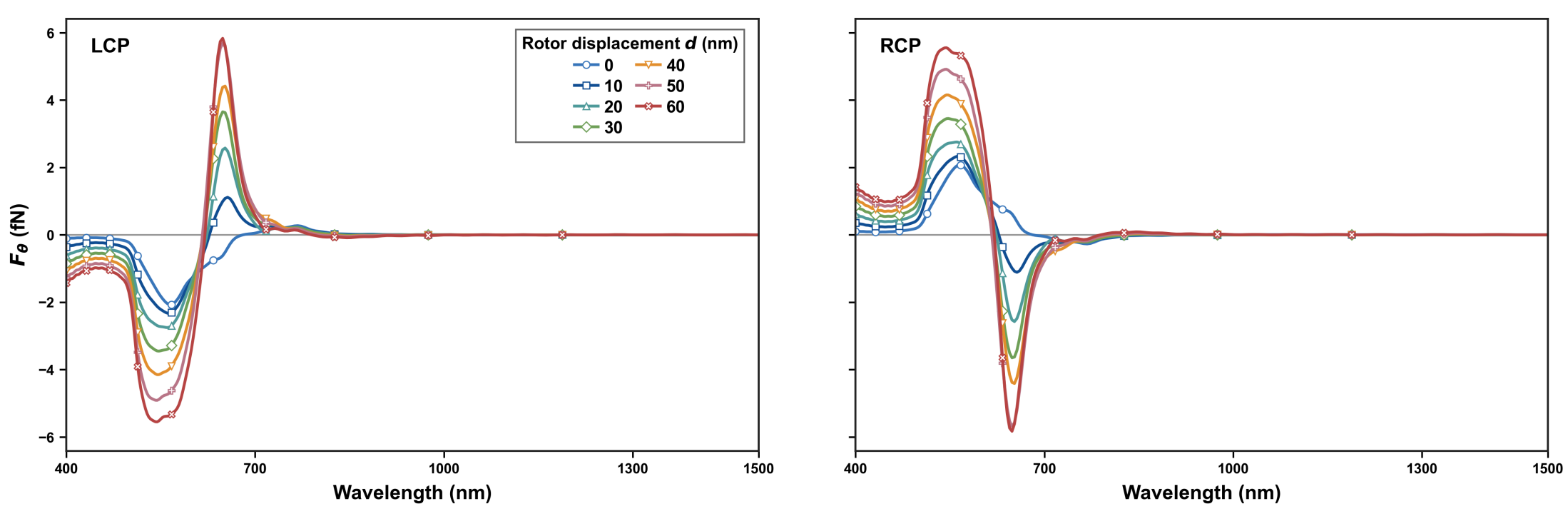

**Figure S7. Helicity-dependent tangential-force response of the opened-shell architecture.** Tangential-force spectra $F_\theta$ under (a) LCP and (b) RCP illumination for rotor displacements $d = 0 - 60\ nm$. The geometry is identical to the opened-shell configuration in Figure 4g of the main text.

### S4.4 Geometric Robustness of the Enclosing Shell

The shell inner radius $R_{\mathrm{in}}$ and thickness $t$ were varied systematically to test the geometric robustness and spectral tunability of the tangential-force response. The corresponding displacement-resolved spectra are shown in Figure S8, and the extracted dominant peak wavelengths are summarized in Figure 4i of the main text.

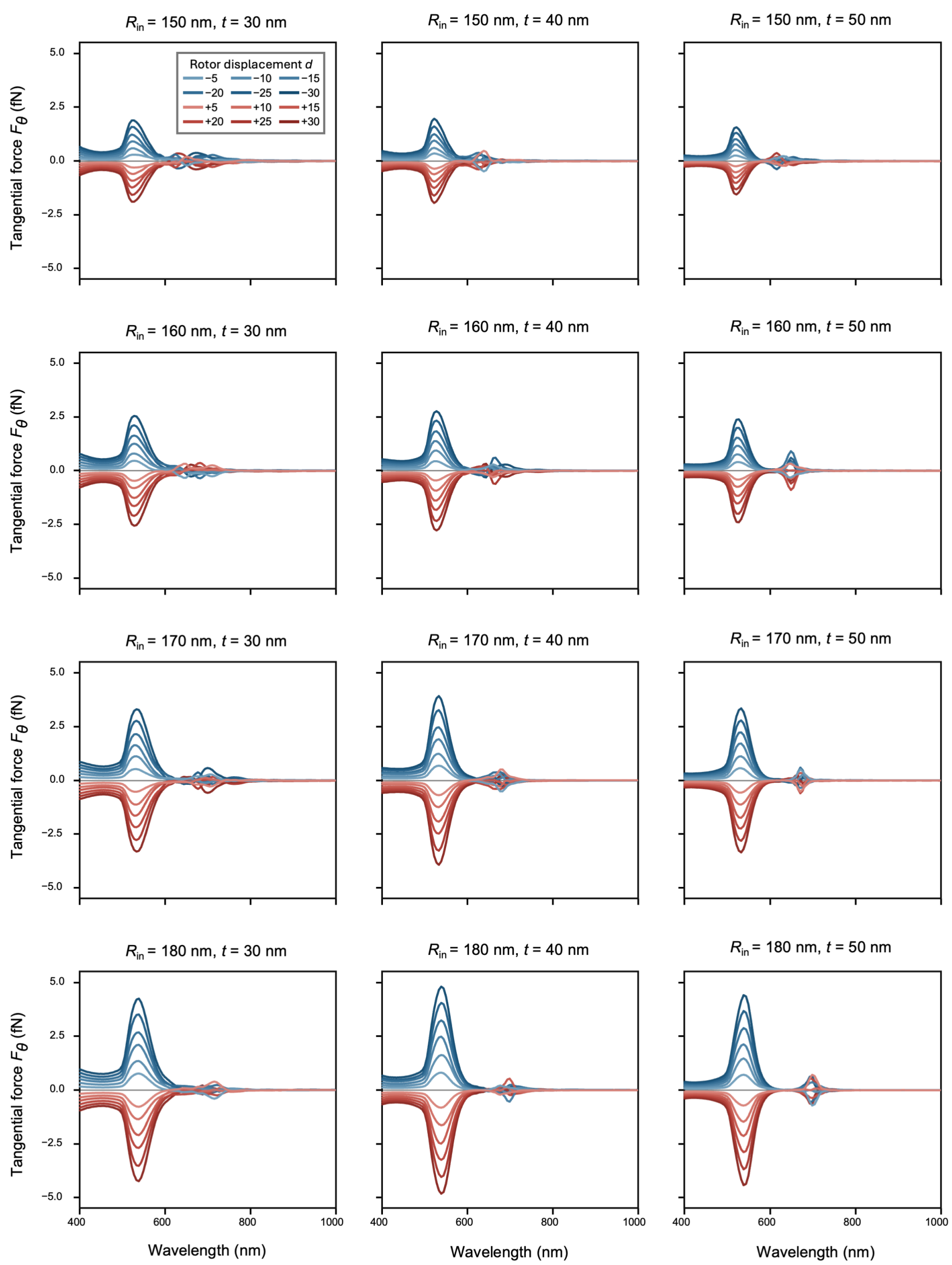

**Figure S8. Tangential-force spectra for varied enclosing-shell geometries.** Displacement-resolved $F_\theta$ spectra for shell stators with inner radii $R_{\text{in}} = 150, 160, 170$, and 180 nm and shell thicknesses $t = 30$, 40, and 50 nm. Curves correspond to the tested signed rotor displacements, with negative and positive displacements shown in blue and red, respectively. The force-peak wavelengths extracted from these spectra are summarized in Figure 4i of the main text.

## Note S5. Material Screening, Bayesian Optimization, and Post-BO Multi-Objective Selection

### S5.1 Material and Rotor-Architecture Screening

A coarse screening of stator–rotor material systems and rotor architectures was first performed to identify a design family combining strong tangential actuation with reduced optical absorption. The enclosing stator had inner and outer radii of 180 and 210 nm, respectively. For each Au-, Ag-, and Si-based material system, the stator and the optically active rotor material were varied together. Solid-sphere rotors had a radius of 60 nm. For the nanoshell architecture, the rotor consisted of a fixed $SiO_2$ core surrounded by a shell whose material matched that of the stator, with inner and outer radii of 40 and 60 nm, respectively. The absorption cross section was calculated as $C_{\text{abs}} = P_{\text{abs}}/I_{\text{inc}}$, where $P_{\text{abs}} = \iiint q_{\text{abs}}(\boldsymbol{r})\text{d}V$ is the total volumetric absorbed power of the stator–rotor structure.

Only tangential-force peaks associated with an outward radial optical force were considered eligible, because the confinement model used below relies on outward optical forcing balanced by inward steric repulsion. The corresponding force and absorption spectra are shown in Figures S9 and S10. Considering the eligible force-to-absorption response together with geometric tunability, the Ag nanoshell was retained for the subsequent geometry search.

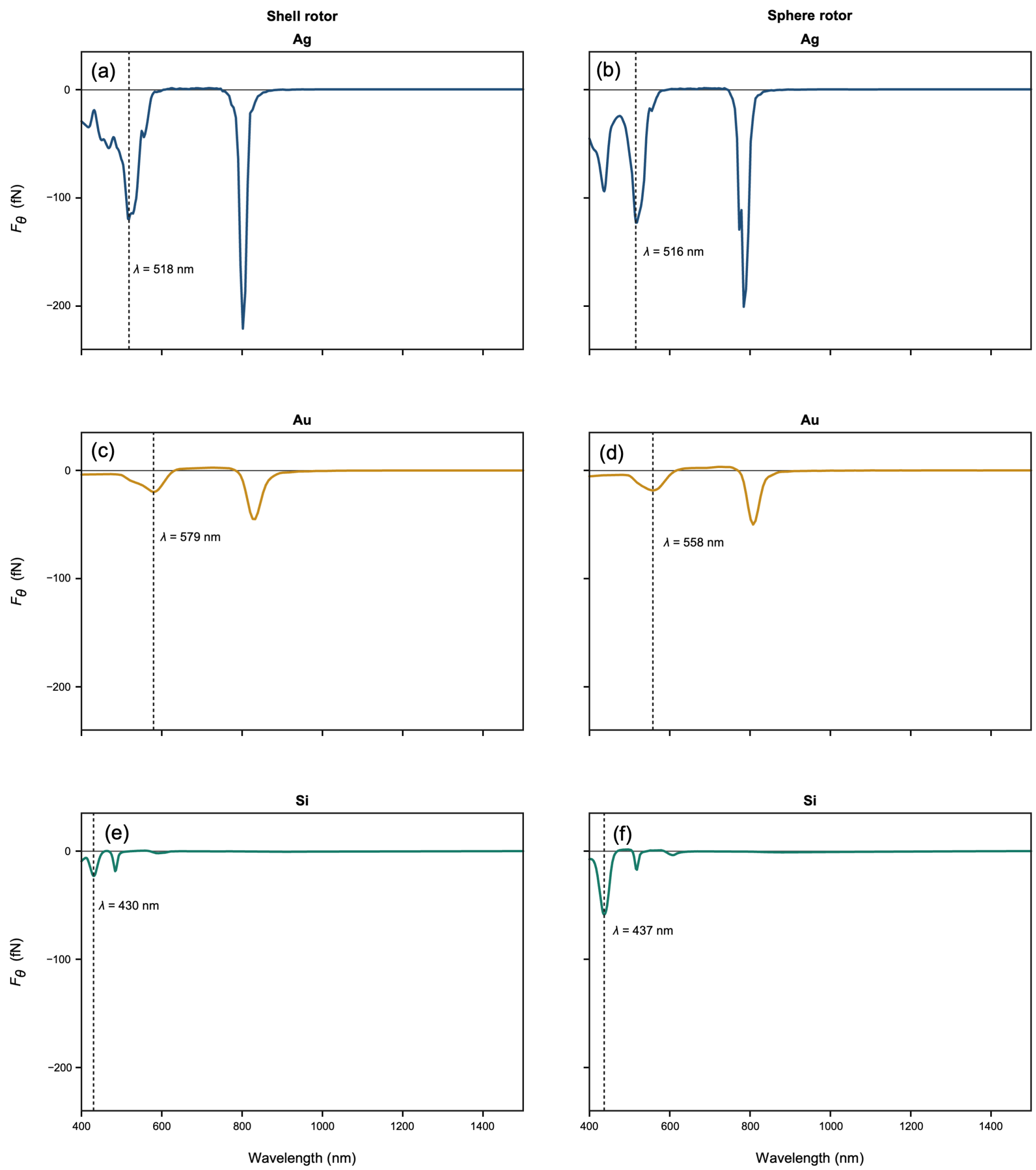

**Figure S9. Tangential-force spectra used for rotor material and architecture screening.** Tangential-force spectra $F_\theta$ for (a,c,e) nanoshell rotors and (b,d,f) solid-sphere rotors composed of (a,b) Ag, (c,d) Au, and (e,f) Si, respectively. Vertical dashed lines mark the eligible force peaks used in the screening comparison. The screening geometries are specified in Section S5.1.

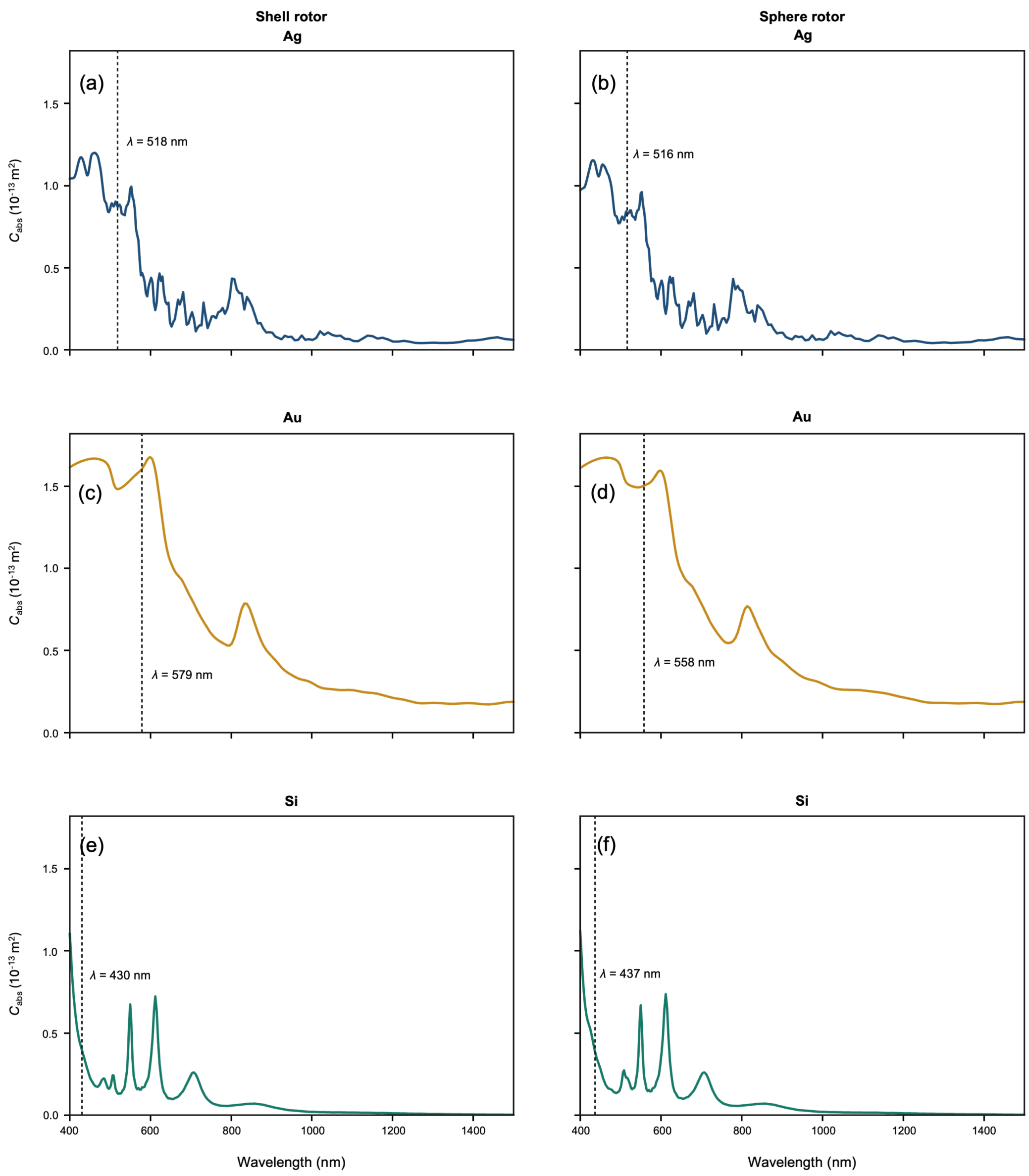

**Figure S10. Absorption spectra used for rotor material and architecture screening.** Absorption cross-section spectra $C_{\mathrm{abs}}$ for (a,c,e) nanoshell rotors and (b,d,f) solid-sphere rotors composed of (a,b) Ag, (c,d) Au, and (e,f) Si. The geometries and illumination conditions are identical to those in Figure S9. Vertical dashed lines indicate the eligible force-peak wavelengths used for the screening comparison.

### S5.2 Bayesian Optimization of the Ag Nanoshell Rotor

The Ag nanoshell was searched in a two-dimensional design space defined by the $SiO_2$-core radius $R_{\mathrm{core}}$ and the Ag-shell thickness $t_{\mathrm{Ag}}$, with the rotor displacement fixed at 60 nm along the +x direction. The search ranges were $15 \leq R_{\mathrm{core}} \leq 35$ nm and $15 \leq t_{\mathrm{Ag}} \leq 40$ nm. The outer radius was defined as

$$R_{\mathrm{out}} = R_{\mathrm{core}} + t_{\mathrm{Ag}}$$

and candidate geometries were retained only when $R_{\mathrm{out}} \leq 60$ nm and $0.30 \leq R_{\mathrm{core}}/R_{\mathrm{out}} \leq 0.70$.

Each geometry was evaluated from its force and absorption spectra over 450–650 nm. The eligible peak wavelength $\lambda_{\mathrm{peak}}$ was defined as the spectral point with the largest $|F_y|$, subject to $F_x > 0$ at that point and its two adjacent spectral samples. At the fixed $+x$ displacement, $F_y$ is the tangential-force component. The optimization objective was

$$J = F_{\mathrm{peak}}/C_{\mathrm{abs}}(\lambda_{\mathrm{peak}}), \quad F_{\mathrm{peak}} = |F_y(\lambda_{\mathrm{peak}})| \tag{S15}$$

The Gaussian-process (GP) surrogate was trained on $\log J$ rather than directly on $J$ to accommodate the dynamic range of the objective. The optimization was initialized with 40 geometrically valid designs generated using a scrambled two-dimensional Sobol sequence[5, 6] with a fixed random seed of 20260718. The two design variables were standardized before fitting a Gaussian-process regression model[7] consisting of a constant kernel multiplied by a Matérn kernel $(\nu = 2.5)$ with an additive white-noise term. The response was normalized internally, and five hyperparameter-optimization restarts were used for each fit. The complete GP kernel is given by

$$k = C \cdot k_{\mathrm{Matérn}}(\nu = 2.5) + k_{\mathrm{White}} \tag{S16}$$

where the constant-kernel bounds were $10^{-3} - 10^{3}$, the Matérn length-scale bounds were $10^{-2} - 10^{2}$, the initial white-noise level was $10^{-6}$ with bounds $10^{-12}$–1, the response was normalized internally, and five hyperparameter-optimization restarts were used.

For each subsequent round, a pool of 20,000 valid Sobol candidates was generated and evaluated by expected improvement (EI) for maximization.[8] With $\mu$ and $\sigma$ denoting the GP predictive mean and standard deviation in log-objective space and $f_{\mathrm{best}}$ the best observed log objective, the acquisition was

$$\mathrm{EI} = (\mu - f_{\mathrm{best}})\Phi(z) + \sigma\phi(z), \; z = (\mu - f_{\mathrm{best}})/\sigma \tag{S17}$$

where $\Phi$ and $\phi$ are the cumulative distribution function and probability density function of the standard normal distribution, respectively. Sixteen candidates were selected greedily from the highest-EI values while enforcing a minimum Euclidean separation of 0.025 in the independently scaled unit-square design space from previously evaluated and newly selected designs. Four additional valid Sobol points were included in each round to maintain broader exploration, giving a batch size of 20. The optimization was terminated after 320 successfully scored geometries.

The geometry with the highest original Bayesian-optimization objective had $R_{\text{core}} = 32.326$ nm and $R_{\text{out}} = 59.906$ nm, corresponding to an Ag-shell thickness of 27.581 nm. This geometry is referred to below as the original BO optimum. It was not automatically adopted as the final device because the original objective primarily targeted tangential actuation rather than radial confinement.

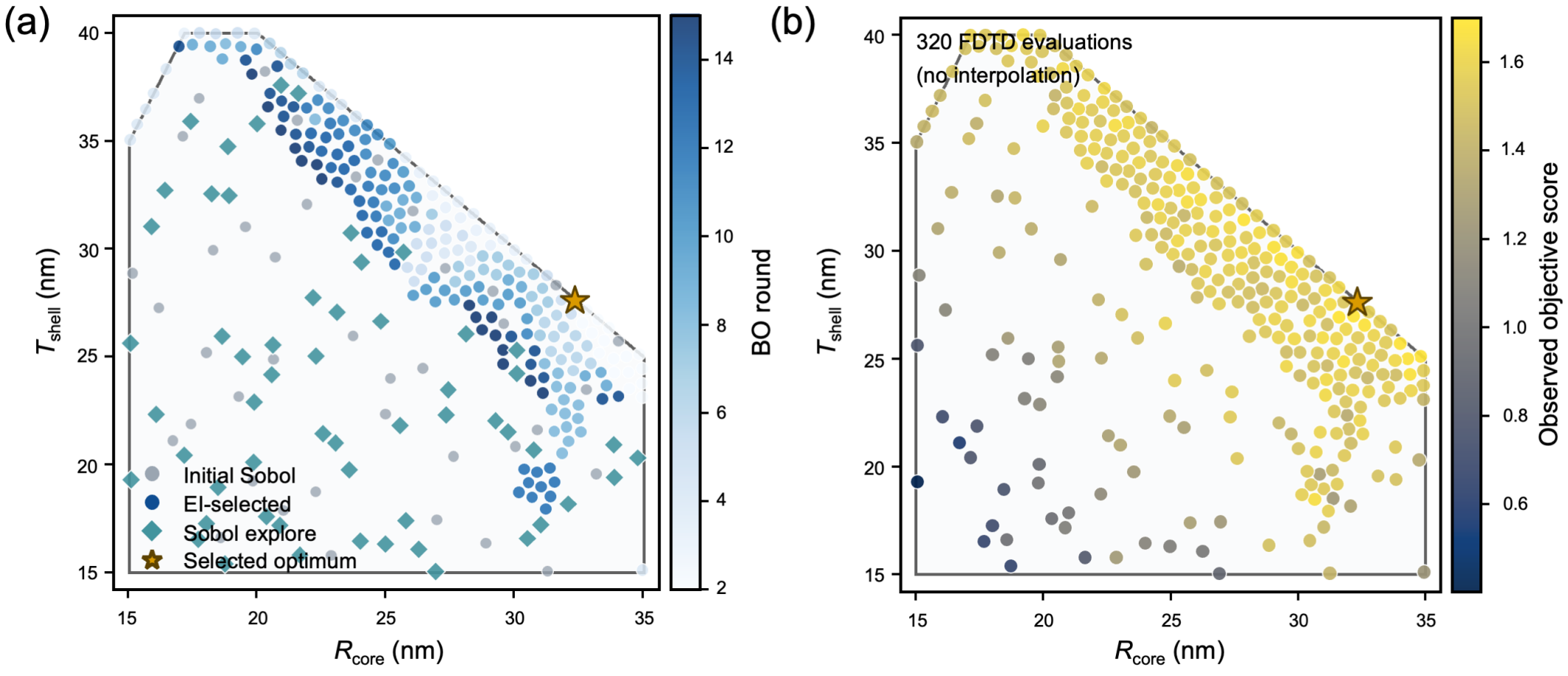

**Figure S11. Sampling and objective values of the original single-objective Bayesian optimization**. (a) Sampled geometries in the constrained $R_{\text{core}}$ $-t_{\text{Ag}}$ design space. (b) Original BO objective for the 320 successfully evaluated geometries. The star marks the geometry with the highest original objective.

### S5.3 Geometry–Wavelength Operating-Point Formulation and Empirical Pareto Analysis

The tangential- and radial-force efficiencies of a given geometry generally peak at different wavelengths. To avoid combining responses that cannot be realized simultaneously, every post-BO operating point was therefore defined by one geometry $g$ at one common wavelength $\lambda$:

$$\eta_\theta = \frac{|F_\theta(g,\ \lambda)|}{C_{\text{abs}}(g,\lambda)}$$

$$\eta_r = \frac{F_r(g,\ \lambda)}{C_{\text{abs}}(g,\lambda)}, \qquad F_r > 0$$

Here, $\eta_r$ is used as an outward radial-force proxy rather than a complete measure of confinement.

Eligible wavelengths were first subjected to nondominated sorting within each geometry, after which the remaining operating points were pooled for cross-geometry sorting. Of 21,593 eligible geometry–wavelength points from the 320 evaluated geometries, 2,402 remained after within-geometry pruning and 20 operating points from 14 geometries formed the empirical Pareto front (Figure S12). This front refers only to the sampled FDTD design library, not the continuous design space.

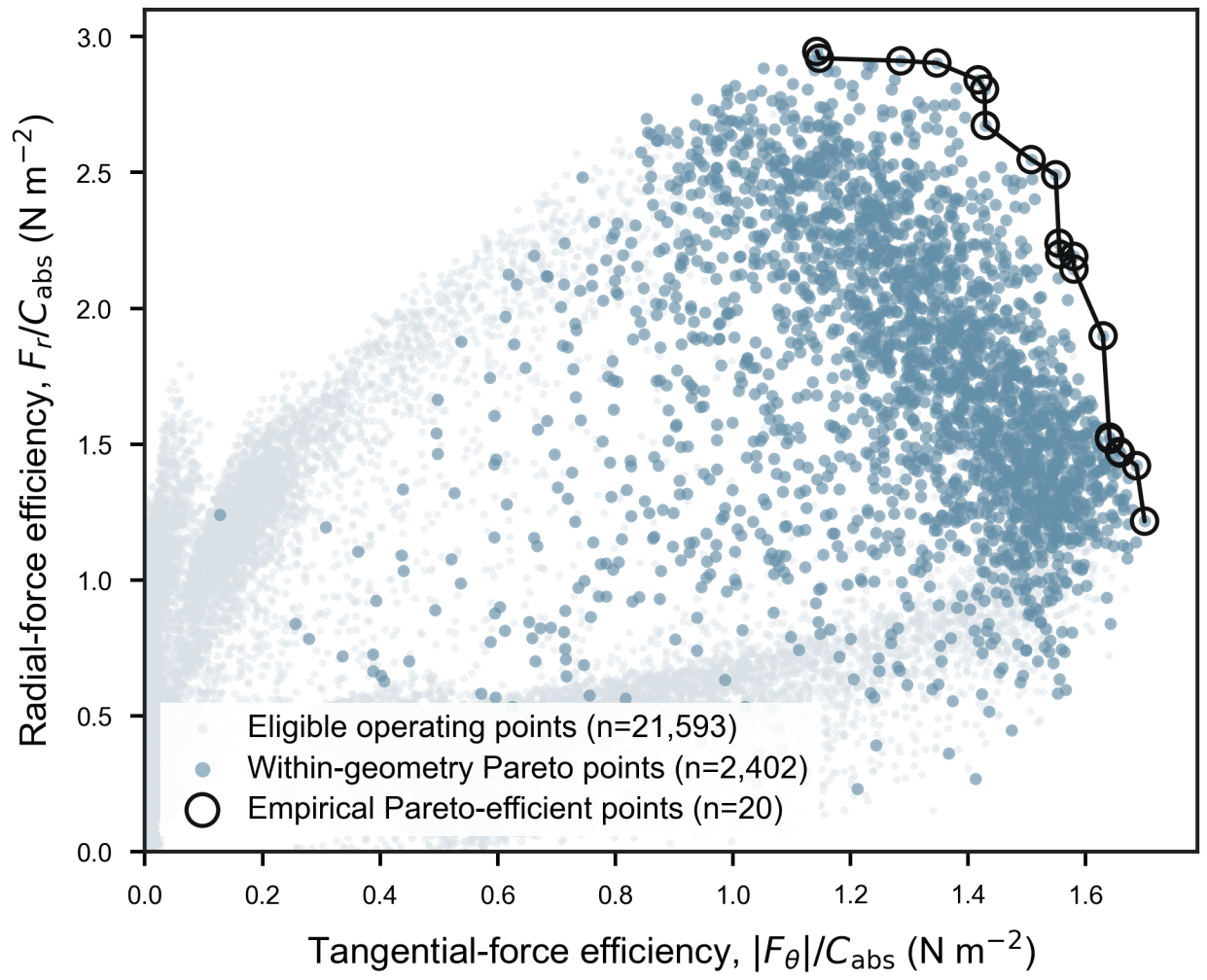

**Figure S12. Geometry–wavelength empirical Pareto analysis.** Light and dark points denote all eligible and within-geometry nondominated operating points, respectively. Open circles connected by the black line mark the empirical Pareto front obtained from the 320 evaluated geometries. Both force efficiencies at each point are evaluated at the same geometry and wavelength.

### S5.4 Selection of a Balanced Pareto Operating Point

To select a balanced operating point without introducing an arbitrary weighted sum, the two Pareto objectives were normalized as

$$X_i = \frac{\eta_{\theta,i}}{\max \eta_\theta}, \qquad Y_i = \frac{\eta_{r,i}}{\max \eta_r},$$

and a maximin score was defined as

$$B_i = \min(X_i, Y_i).$$

The maximum $B_i$ corresponds to Candidate 02 (C2), with $R_{\text{core}} = 33.3\ nm, t_{\text{Ag}} = 25.2\ nm, and\ R_{\text{out}} = 58.5\ nm$, operated at $\lambda = 534\ nm$.

For C2, the independent tangential- and radial-efficiency maxima occur at 528.552 and 536.388 nm, respectively. At the selected wavelength, 96.45% and 90.63% of these respective maxima are retained (Figure S13).

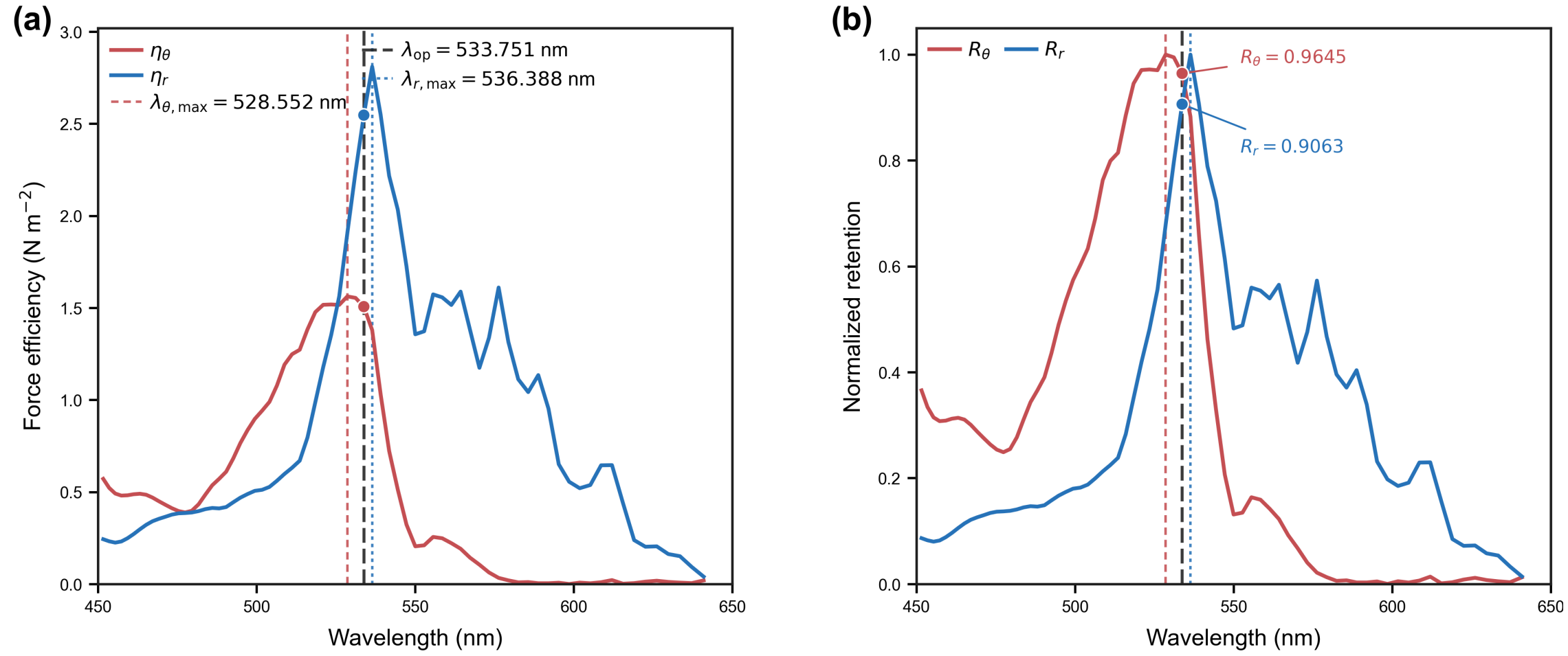

**Figure S13. Spectral response of Candidate 02.** (a) Tangential-force efficiency $\eta_\theta$ and outward radial-force efficiency $\eta_r$ of Candidate 02 as functions of wavelength. Vertical lines mark their independent maxima and the selected operating wavelength. (b) Corresponding efficiencies normalized by their respective spectral maxima.

## Note S6. Displacement-Resolved Mechanical Validation of Pareto Candidates

### S6.1 Displacement-Resolved Optical-Force Profiles

Five representative operating points were selected to test whether the spectral trade-off identified in Section S5 persists in the displacement-dependent optical forces relevant to orbital dynamics (Table S1). Each candidate was evaluated at its fixed operating wavelength.

For each candidate, the radial and tangential optical forces, $F_r(r)$ and $F_\theta(r)$, were calculated by FDTD/MST over rotor displacements from 0 to 75 nm, with additional sampling near the radial force-balance region. Shape-preserving piecewise cubic Hermite interpolation was used only within the sampled displacement range (Figure S14).[9]

**Table S1. Representative Pareto candidates used for displacement-resolved validation**.

| Candidate | Role | $R_{\text{core}}$(nm) | $t_{\text{Ag}}$(nm) | $R_{\text{out}}$(nm) | $\lambda_{\text{op}}$(nm) |
|---|---|---|---|---|---|
| C1 | Radial endpoint | 26.775 | 31.595 | 58.369 | 536.388 |
| C2 | Balanced Pareto candidate | 33.268 | 25.154 | 58.422 | 533.751 |
| C3 | Original BO optimum | 32.326 | 27.581 | 59.906 | 518.454 |
| C4 | Intermediate Pareto candidate | 30.649 | 27.775 | 58.424 | 533.751 |
| C5 | Intermediate Pareto candidate | 33.224 | 26.717 | 59.941 | 533.751 |

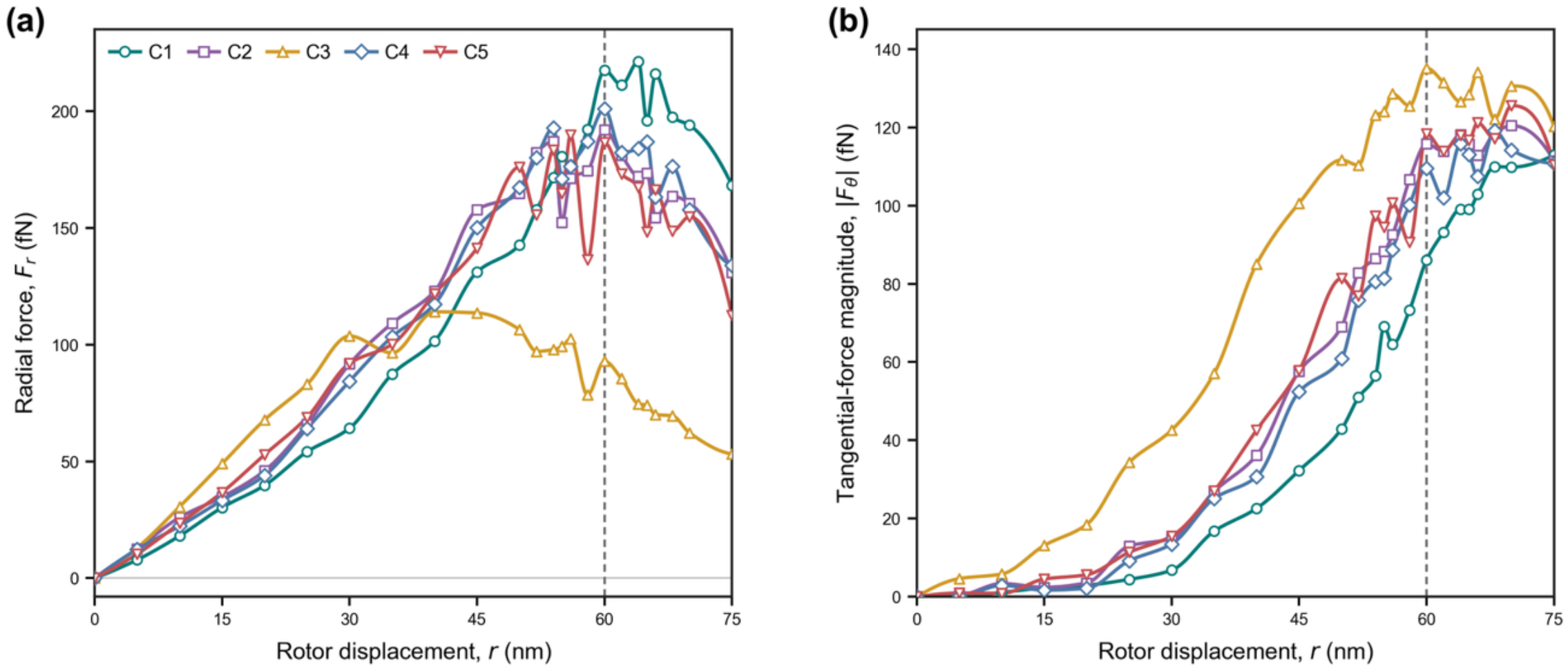

**Figure S14. Displacement-resolved optical-force profiles of representative Pareto candidates.** (a) Radial force $F_r$ and (b) tangential force $F_\theta$ as functions of rotor displacement. Symbols denote direct FDTD/MST calculations and solid curves denote shape-preserving interpolants within the sampled 0–75 nm displacement range. The dashed line marks the nominal soft-wall position $R_w = 60\ nm$.

### S6.2 Radial Force Balance and Effective Radial Potential

To compare the radial confinement supported by the five candidates, the optical radial force was combined with the same short-range PEG-like soft-wall interaction,

$$F_{\mathrm{PEG}}(r) = -F_0 e^{\frac{r-R_w}{\lambda_{\mathrm{PEG}}}}$$

where $R_w = 60\ \mathrm{nm}$ denotes the nominal soft-wall position. For the deterministic comparison in this section, $F_0 = 0.40\ \mathrm{pN}$ and $\lambda_{\mathrm{PEG}} = 0.50\ \mathrm{nm}$ were used for all candidates. The total radial force was therefore

$$F_r^{\mathrm{net}} = F_r^{\mathrm{optical}} + F_r^{\mathrm{PEG}}$$

Stable radial equilibria were identified from

$$F_r^{\mathrm{net}}(r_{\mathrm{eq}}) = 0$$

with $\frac{\mathrm{d}F_r^{\mathrm{net}}}{\mathrm{d}r} < 0$. The corresponding local radial stiffness was evaluated as

$$k_{r,\mathrm{eff}} = -\frac{\mathrm{d}F_r^{\mathrm{net}}}{\mathrm{d}r}|_{r=r_{\mathrm{eq}}}$$

and the effective radial potential was obtained from

$$U_r(r) = -\int F_r^{\mathrm{net}}(r)\,\mathrm{d}r,$$

with $U_r(r_{\mathrm{eq}})$ taken as the reference zero.

All five candidates exhibit a stable radial force balance close to the nominal soft-wall position (Figure S15). The equilibrium radii range from 59.24 to 59.69 nm. The corresponding local radial stiffnesses vary substantially, from 166 fN nm$^{-1}$ for the original BO optimum C3 to 425 fN nm$^{-1}$ for the radial endpoint C1, while the balanced candidate C2 retains a comparatively large stiffness of 374 fN nm$^{-1}$. The resulting effective potentials exhibit local minima at the respective force-balance positions, consistent with restoring radial confinement near the outer orbital region.

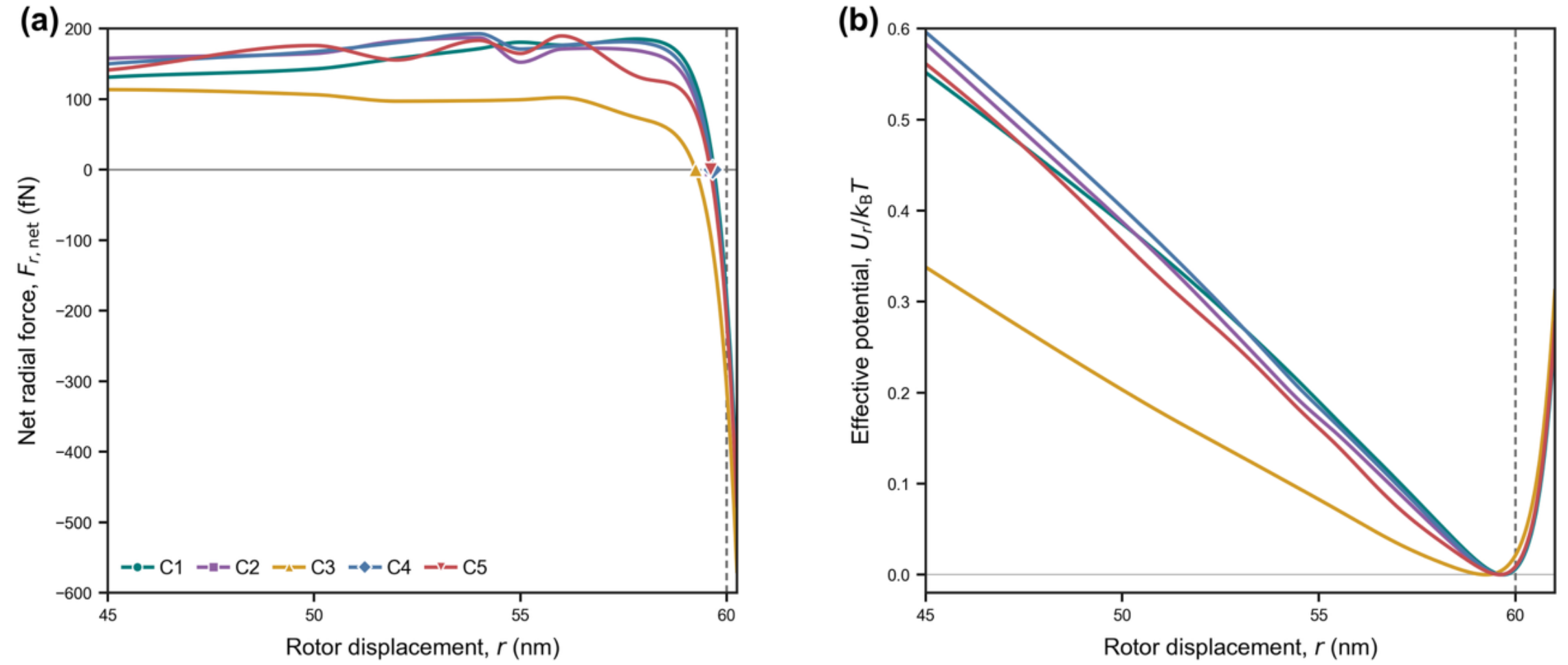

**Figure S15**. **Radial force balance and effective radial potential of representative Pareto candidates.** (a) Net radial force under the common nominal PEG-like soft-wall interaction; markers indicate stable force-balance positions. (b) Corresponding effective radial potentials referenced to zero at each stable equilibrium. The dashed line marks $R_w = 60\ nm$.

**Table S2. Radial force-balance properties under the common nominal soft-wall model**.

| Candidate | $r_{\mathrm{eq}}$(nm) | $k_{r,\mathrm{eff}}$(fN nm$^{-1}$) |
|---|---:|---:|
| C1 | 59.693 | 424.99 |
| C2 | 59.629 | 374.01 |
| C3 | 59.245 | 166.36 |
| C4 | 59.654 | 395.99 |
| C5 | 59.605 | 339.04 |

$R_w = 60\ nm$, $F_0 = 0.40$ pN, and $\lambda_{\mathrm{PEG}} = 0.50\ nm$ were used for all candidates

## Note S7. Photothermal Model and Numerical Validation

### S7.1 Photothermal Model and Numerical Implementation

For the photothermal calculation, the electromagnetic field was evaluated directly at the target incident intensity of $1 \times 10^9$ W m$^{-2}$ in water ($n = 1.333$), using the same source normalization as the corresponding force calculation. The local volumetric absorbed-power density was computed from the electric field as[10]

$$q_{\mathrm{abs}}(\boldsymbol{r}) = \frac{1}{2}\omega\varepsilon_0 \mathrm{Im}[\varepsilon(\boldsymbol{r})]|\boldsymbol{E}(\boldsymbol{r})|^2 \tag{S18}$$

with $|\boldsymbol{E}|^2 = |E_x|^2 + |E_y|^2 + |E_z|^2$. Numerical round-off values below zero were set to zero before thermal interpolation; no $C_{\mathrm{abs}}I$ rescaling was applied.

The steady-state temperature rise ΔT was obtained from

$$\nabla \cdot [\kappa(\boldsymbol{r})\nabla \Delta T(\boldsymbol{r})] + q_{\mathrm{abs}}(\boldsymbol{r}) = 0 \tag{S19}$$

using a finite-volume discretization on a uniform 5 nm grid. The cubic computational domain extended from $-1.6$ to $+1.6$ μm along each Cartesian coordinate. Material-specific bulk thermal conductivities were assigned as $\kappa_{\mathrm{water}} = 0.606$ W m$^{-1}$ K$^{-1}$,[11] $\kappa_{\mathrm{Au}} = 317$ W m$^{-1}$ K$^{-1}$, $\kappa_{\mathrm{Ag}} = 429$ W m$^{-1}$ K$^{-1}$,[12] and $\kappa_{\mathrm{SiO_2}} = 1.38$ W m$^{-1}$ K$^{-1}$.[13] Harmonic averaging of $\kappa$ was used across adjacent finite-volume cells at material interfaces.

A Dirichlet boundary condition $\Delta T = 0$ was imposed on all six outer faces. The discretized linear system was solved using a diagonally preconditioned conjugate-gradient algorithm with a relative tolerance of $10^{-6}$ and a maximum of 12,000 iterations. The maximum steady-state temperature rise was extracted from the converged three-dimensional field. For the final selected Ag-nanoshell case, the thermal source was evaluated at $\lambda = 534$ nm, with a $SiO_2$-core radius of 33.3 nm, an outer Ag radius of 58.4 nm, and an in-plane rotor offset of 60 nm.

### S7.2 Finite-Volume Heat-Solver Validation and Grid Sensitivity

The custom finite-volume (FV) thermal solver was validated against an analytical benchmark before being applied to the nanorotor geometries. The benchmark consisted of a one-dimensional heterogeneous heat-conduction problem containing a water–Ag interface, for which the steady-state temperature distribution and the interfacial heat flux can be evaluated analytically. The same finite-volume discretization and harmonic-mean treatment of the thermal conductivity at material interfaces as those used in the three-dimensional photothermal calculations were retained in the benchmark. At the production grid spacing of 5 nm, the numerical temperature profile closely reproduces the analytical solution (Figure S16a,b).

Grid convergence of the benchmark is shown in Figure S16c. Both the relative $L_2$ temperature error and the interfacial heat-flux error decrease systematically with grid refinement, with the temperature error following approximately second-order convergence. At a 5 nm grid spacing, the relative temperature-field error is approximately 0.1%.

Grid sensitivity was also evaluated directly for the optimized Ag-nanoshell geometry using the same FDTD-derived heat source (Figure S16d). Relative to the 5 nm result, the rotor-center temperature varies by less than 1% over the tested grid spacings, while the deviation in the maximum temperature rise remains below approximately 4.1% at the coarsest tested grid.

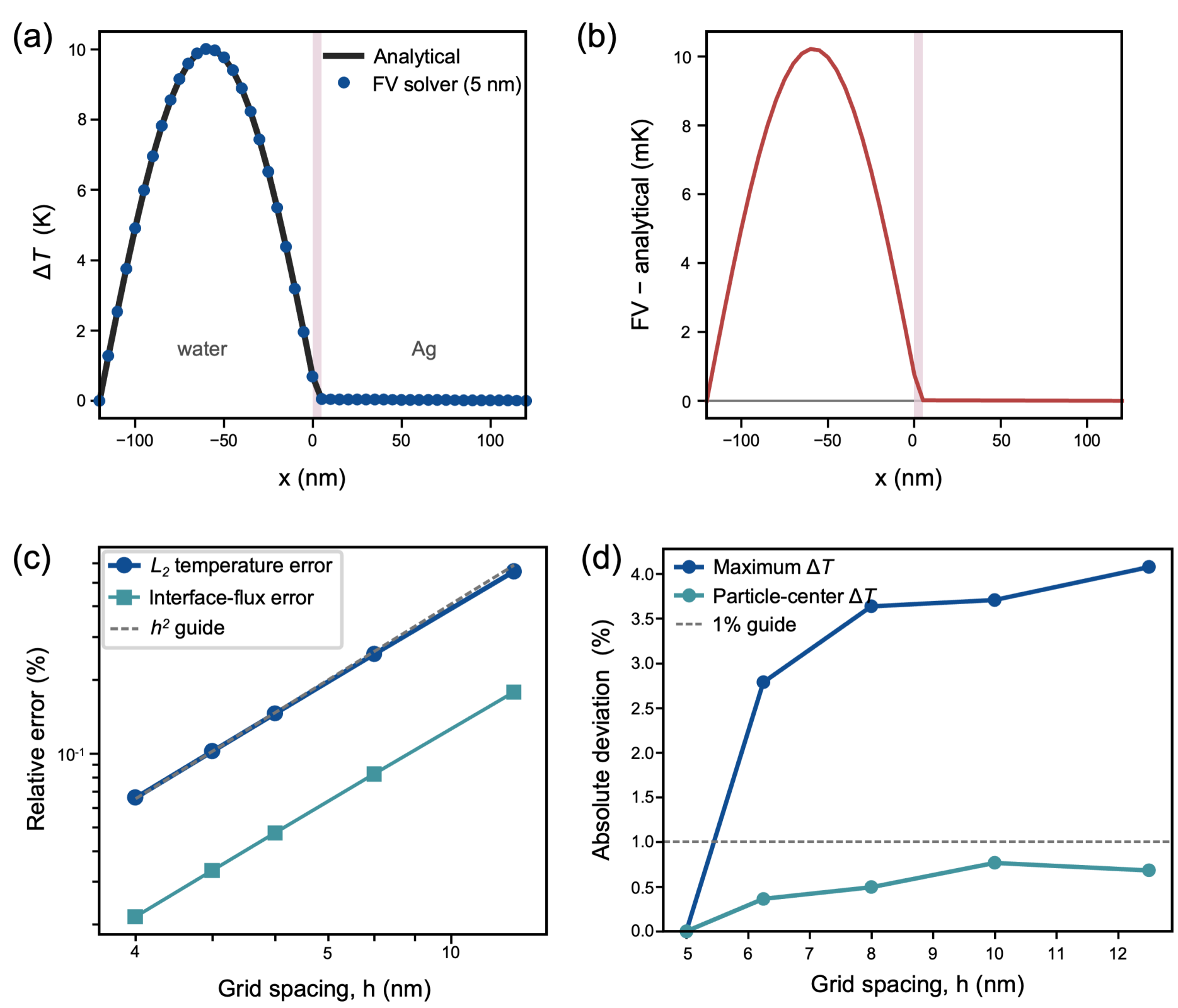

**Figure S16. Validation and grid sensitivity of the finite-volume photothermal model.** (a) Analytical and finite-volume (FV) steady-state temperature profiles for a one-dimensional water–Ag benchmark at a 5 nm grid spacing. (b) Pointwise temperature difference for the same benchmark. (c) Relative $L_2$ temperature-field error and relative interfacial heat-flux error as functions of grid spacing; the dashed line indicates an $h^2$ scaling guide. (d) Grid sensitivity of the final selected Ag-nanoshell calculation relative to the 5 nm result.

## Note S8. FDT-Consistent Brownian Dynamics and Ensemble Validation

### S8.1 FDT-Consistent Stochastic Model

The projected orbital dynamics were modeled using a two-dimensional overdamped Langevin equation.[14] The deterministic in-plane force was

$$\boldsymbol{F}(\boldsymbol{r}) = F_r^{\mathrm{net}}(\boldsymbol{r})\hat{\boldsymbol{e}}_r + F_\theta(\boldsymbol{r})\hat{\boldsymbol{e}}_\theta, \tag{S20}$$

where $F_r^{\mathrm{net}} = F_r^{\mathrm{optical}} + F_r^{\mathrm{PEG}}$ includes the optical radial force and the short-range PEG-like soft-wall interaction defined in Section S6.2. The stochastic dynamics were integrated according to

$$\mathrm{d}\boldsymbol{r} = \mu\boldsymbol{F}(\boldsymbol{r})\mathrm{d}t + \sqrt{2D}\mathrm{d}\boldsymbol{W}, \tag{S21}$$

with the translational mobility and diffusion coefficient related by

$$\mu = \frac{1}{6\pi\eta a},\ D = \mu k_{\mathrm{B}}T, \tag{S22}$$

Here, $a$ is the outer radius of the corresponding nanoshell rotor and $\mathrm{d}\boldsymbol{W}$ is a two-dimensional Wiener increment. Thus, the stochastic noise and deterministic mobility satisfy the fluctuation–dissipation relation without an independent noise-scaling parameter.

Candidate-specific optical-force profiles were interpolated by shape-preserving piecewise cubic Hermite interpolation within the directly sampled 0–75 nm displacement range. Trajectories leaving this range were classified as escaped; no optical-force extrapolation, force freezing, or reflecting boundary was used. Simulation parameters are summarized in Table S3.

**Table S3**. **Physical and numerical parameters used in the FDT-consistent Brownian-dynamics simulations**.

| **Parameter** | **Value** |
|---|---|
| $T$ | 298.15 K |
| $\eta$ [15] | $0.890 \times 10^{-3}$ Pa s |
| Hydrodynamic radius $a$ | candidate-specific $R_{\mathrm{out}}$ |
| $R_w$ | 60 nm |
| $\lambda_{\mathrm{PEG}}$ | 0.40–0.60 nm |
| $F_0$ | 0.32–0.48 pN |
| $\Delta t$ | 0.5 μs |
| Duration | 1 s |
| Steady-state window | 0.1–1.0 s |
| Trajectories/condition | 32 |
| Force domain | 0–75 nm |

The hydrodynamic radius was taken as the actual outer radius of each candidate nanoshell. The PEG parameter ranges were used for the systematic screening described in Section S8.2.

### S8.2 Candidate and PEG-Parameter Screening

To assess the dependence of the stochastic dynamics on the phenomenological soft-wall parameters, each of the five candidates in Table S1 was evaluated over a predefined $5 \times 5$ PEG-parameter grid,

$$\lambda_{\mathrm{PEG}} = \{0.40, 0.45, 0.50, 0.55, 0.60\}\ \mathrm{nm},$$

$$F_0 = \{0.32, 0.36, 0.40, 0.44, 0.48\}\ \mathrm{pN}$$

For every candidate–parameter combination, 32 independent 1 s trajectories were simulated, giving 4,000 trajectories in total. Radial confinement was characterized by the central-90% radial width, $r_{95} - r_5$, whereas directional propulsion was characterized by the magnitude of the ensemble-averaged angular velocity, $|\langle w \rangle|$. Escape fraction and the fraction of trajectories retaining the helicity-defined circulation direction were also monitored. Across all 125 tested parameter sets, no trajectory escaped the directly sampled force domain and all trajectories retained the same circulation direction. The practical parameter dependence within the tested range therefore primarily reflects a trade-off between radial-distribution width and circulation rate (Figure S17).

The confinement-oriented comparison gave the numerically smallest central-90% radial width for C2 (40.686 nm), although C4 and C5 showed closely comparable values. By contrast, the original BO optimum C3 provided the largest circulation rate, whereas C2 retained the second-highest candidate-level circulation rate. These results are consistent with the distinct radial–tangential roles identified from the displacement-resolved optical-force profiles and support C2 as a balanced rather than single-metric optimum.

For the subsequent ensemble analysis, the C2 parameter sets were further compared without an explicit weighted sum. The normalized maximin point on its parameter-level confinement–circulation trade-off corresponded to

$$\lambda_{\mathrm{PEG}} = 0.55\ \mathrm{nm}, \qquad F_0 = 0.48\ \mathrm{pN}.$$

This condition gives a central-90% radial width of 40.793 nm and a mean angular velocity of $1.350 \times 10^3$ rad $\mathrm{s}^{-1}$, while retaining zero escape and 32/32 helicity-consistent trajectories. It was therefore used for the final ensemble statistics described in Section S8.3.

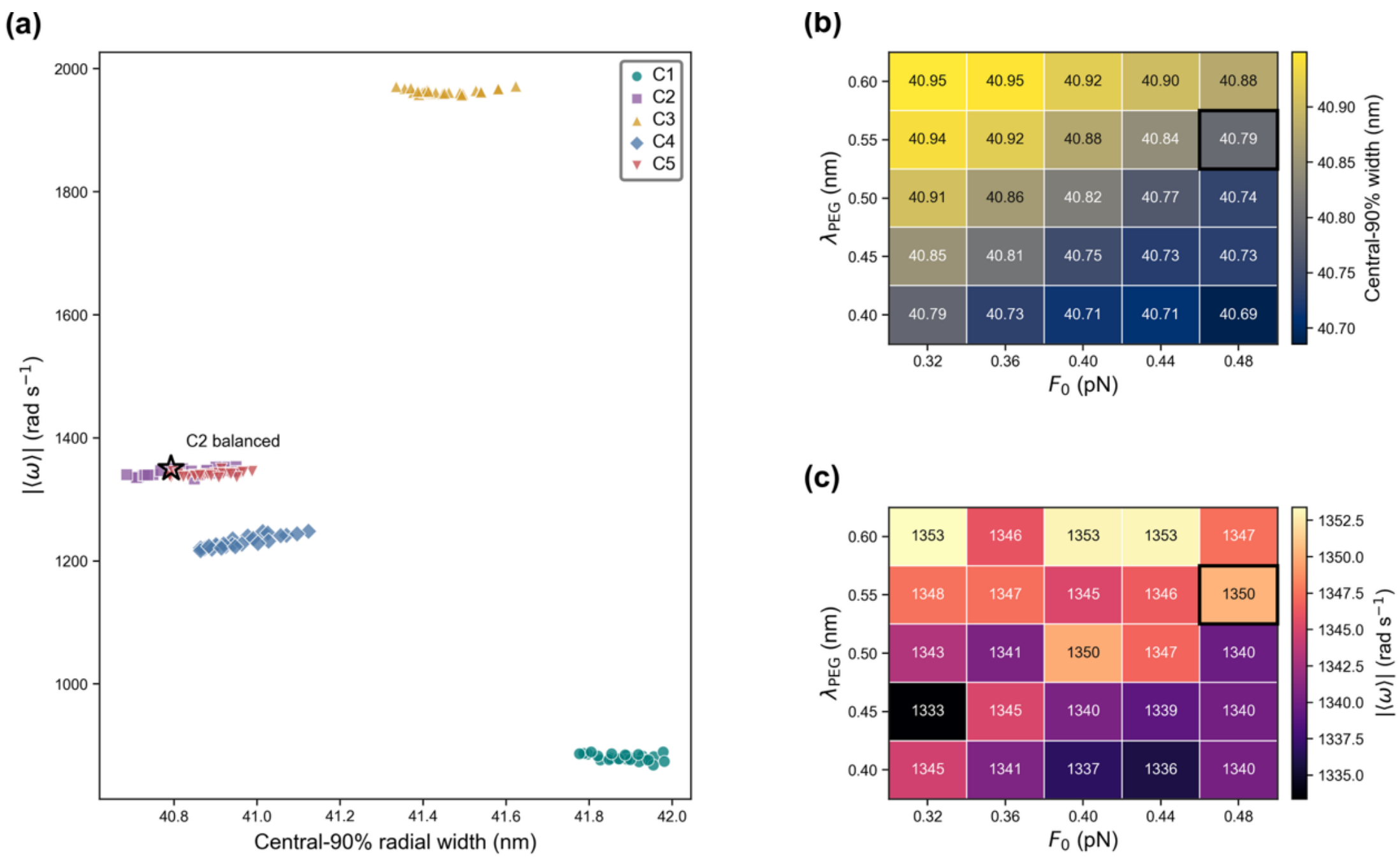

**Figure S17**. **Brownian-dynamics screening across candidate devices and PEG-wall parameters**. (a) Central-90% radial width versus angular speed for the 125 candidate–parameter combinations; each point represents 32 independent trajectories. (b,c) Central-90% radial width and angular speed, respectively, for C2 over the $5 \times 5$ PEG-parameter grid. The highlighted condition was used for the final ensemble analysis.

### S8.3 Final Ensemble Statistics for Candidate 02

The final C2 ensemble used $\lambda_{\mathrm{PEG}} = 0.55\,\mathrm{nm}$ and $F_0 = 0.48\,\mathrm{pN}$. All 32 independent trajectories were retained, and steady-state spatial statistics were evaluated over $t = 0.1 - 1.0$s. The radial distribution had a median of 50.667 nm, with $r_5 = 19.932\,\mathrm{nm}$ and $r_{95} = 60.725$, respectively, corresponding to a central-90% width of 40.793 nm (Figure S18a). The two-dimensional occupancy remains strongly weighted toward an outer annulus adjacent to the soft wall, while extending substantially inward under thermal fluctuations (Figure S18b). No trajectory escaped the sampled force domain.

The accumulated angular displacement increases persistently in the helicity-selected circulation direction for all trajectories (Figure S18c). The magnitude of the ensemble-averaged angular velocity was

$$\langle w \rangle = 1.350 \times 10^3 \text{ rad s}^{-1},$$

with a trajectory-level bootstrap 95% confidence interval of

$$[1.302, 1.395] \times 10^3 \text{ rad s}^{-1}.$$

All 32 trajectories were helicity-consistent in the corresponding signed analysis (Figure S18d).

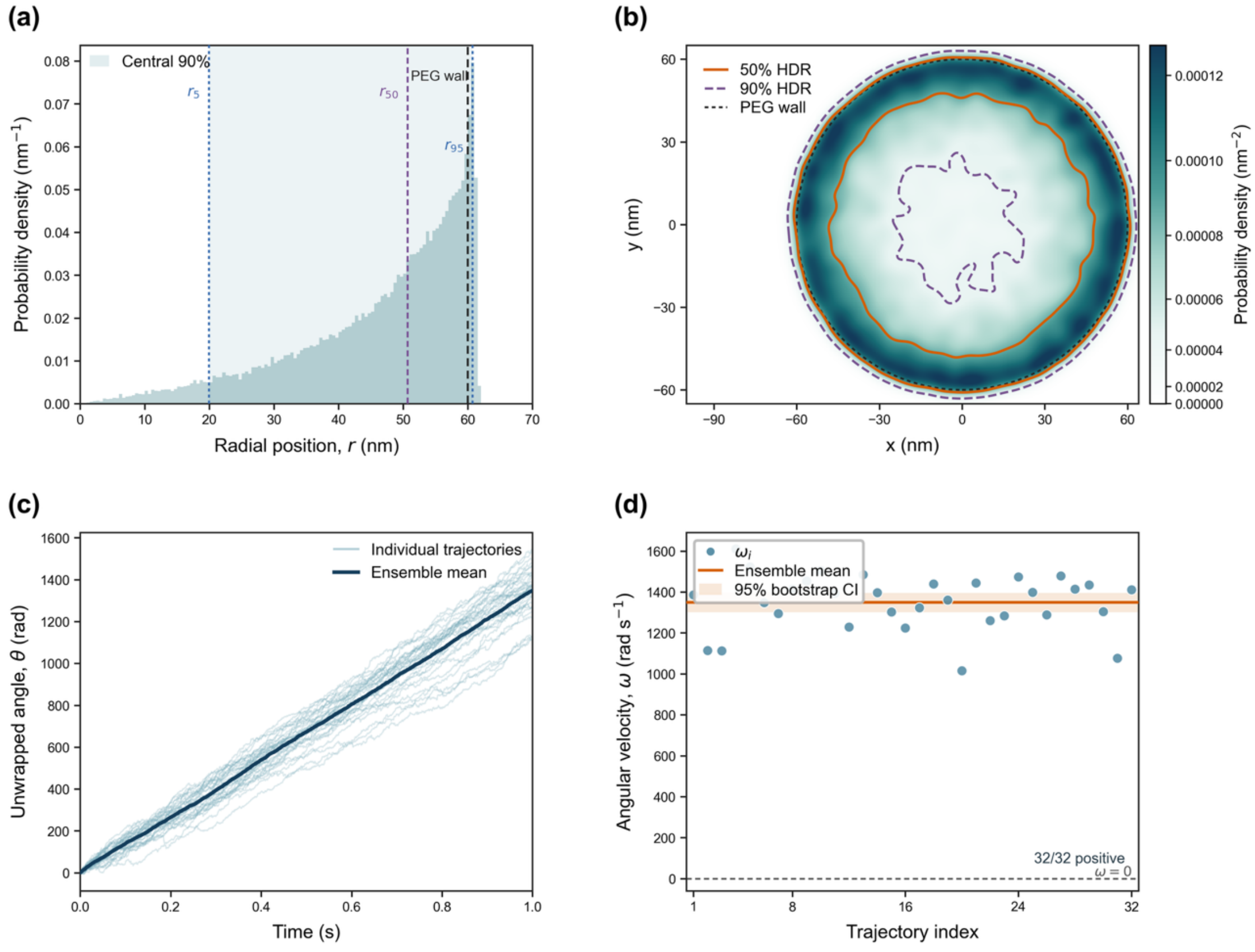

**Figure S18**. **Ensemble Brownian statistics of the final Candidate 02 device**. (a) Steady-state radial probability distribution obtained from all 32 trajectories over $t = 0.1 - 1.0$s. The dashed markers indicate the 5th, median, and 95th percentiles at 19.932, 50.667, and 60.725 nm, respectively; the vertical dashed line marks the nominal PEG soft-wall position at 60 nm. (b) Corresponding two-dimensional steady-state occupancy density. The inner and outer contours enclose the 50% and 90% highest-density regions,[16] respectively, and the dashed circle denotes the 60 nm PEG wall. (c) Unwrapped angular displacement $\theta_i(t)$ for all 32 independent 1 s trajectories. The thick curve denotes the ensemble mean. (d) Trajectory-level angular velocities $\omega_i$. The horizontal line and shaded interval denote the ensemble mean, $1.350 \times 10^3$ rad s$^{-1}$, and its bootstrap 95% confidence interval,[17] $[1.302, 1.395] \times 10^3$ rad s$^{-1}$, respectively. All trajectories retain the same helicity-defined circulation direction.

### S8.4 Time-Step Convergence

The numerical time-step dependence of the Brownian-dynamics simulations was evaluated using the final C2 condition identified in Section S8.2, with $\lambda_{\mathrm{PEG}} = 0.55$ nm *and* $F_0 =$ 0.48 pN. Simulations were repeated using

$$\Delta t = 0.25, 0.50, 1.00\ \mu\mathrm{s},$$

while keeping the optical-force inputs, temperature, viscosity, confinement parameters, trajectory duration, and ensemble size unchanged. For each time step, 32 independent 1 s trajectories were generated and analyzed over the same steady-state interval of $t = 0.1 - 1.0$s.

Convergence was assessed using three quantities that characterize the final stochastic state: the magnitude of the ensemble-averaged angular velocity, $|\langle w \rangle|$, the central-90% radial width, $r_{95} - r_5$, and the median radial position. Escape events and helicity-consistent circulation were additionally monitored for all trajectories. The results are summarized in Figure S19.

The results obtained with $\Delta t = 0.50$ μs were closely consistent with those obtained using the finer 0.25 μs reference. The magnitude of the ensemble-averaged angular velocity differed by 1.43%, while the central-90% radial width differed by only 0.15%. The median radial position changed by 0.25 nm. Both time steps yielded zero escape and helicity-consistent circulation for all 32 trajectories. By comparison, increasing the time step to 1.00 μs produced a larger 4.02% deviation in angular speed and a 0.58 nm shift in the median radial position. A time step of 0.50 μs was therefore retained for the production simulations.

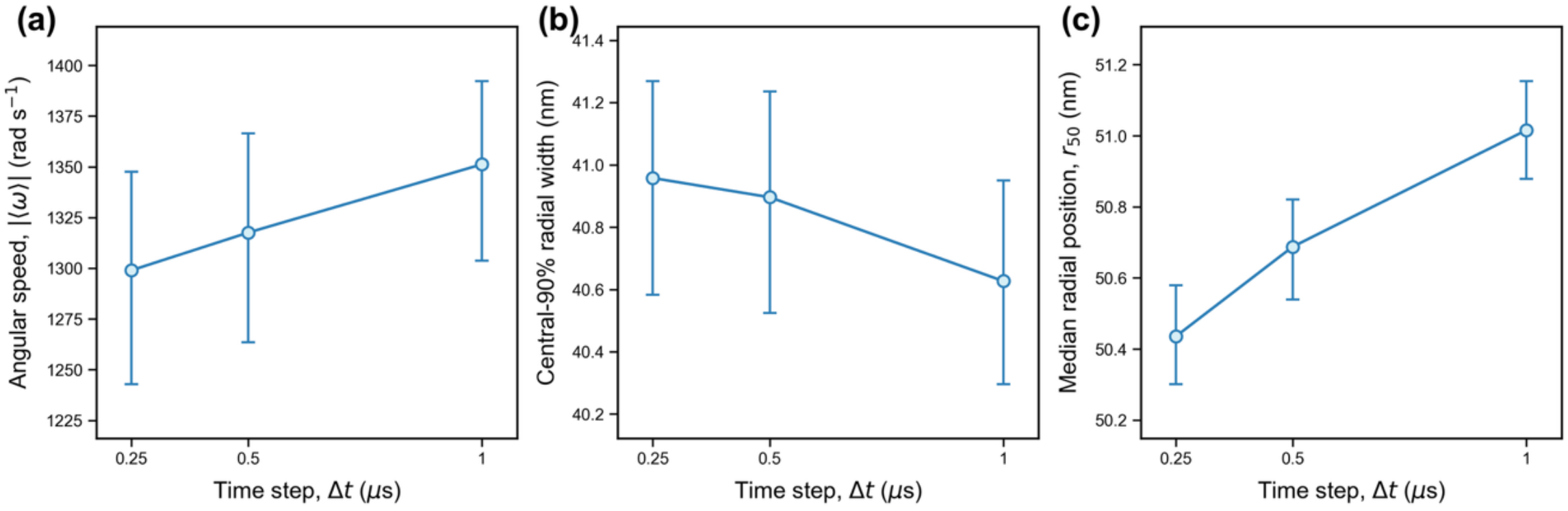

**Figure S19. Time-step convergence of the FDT-consistent Brownian-dynamics simulations. (a)** Magnitude of the ensemble-averaged angular velocity, **(b)** central-90% radial width, and **(c)** median radial position for the final Candidate 02 condition simulated using time steps of 0.25, 0.50, 1.00 μs. Each condition comprises 32 independent 1 s trajectories analyzed over $t = 0.1 - 1.0$ s. Error bars denote 95% confidence intervals obtained by whole-trajectory bootstrap resampling.

### S8.5 Incident-Intensity Dependence of Orbital Dynamics

The simulations above use $1 \times 10^9\ W\ m^{-2}$ as a common reference intensity rather than a unique operating condition. To examine the intensity dependence of the orbital dynamics, the displacement-resolved optical-force profiles of C2 were scaled linearly with incident intensity while keeping the geometry, operating wavelength, PEG-wall parameters, and thermal-bath temperature fixed. Brownian dynamics were evaluated from 32 paired trajectories at each intensity using the same FDT-consistent model as in Sections S8.1–S8.4.

Increasing the incident intensity progressively narrows the stochastic radial distribution and increases the orbital circulation rate (Figure S20a,b). The central-90% radial width decreases monotonically over the tested range, while the angular speed increases approximately linearly. The optical work per cycle relative to thermal energy, $W_{\text{cycle}}/k_{\text{B}}T$, increases correspondingly, indicating a growing deterministic orbital drive relative to thermal fluctuations (Figure S20c). This improvement is accompanied by an approximately linear increase in the steady-state temperature rise (Figure S20d). Thus, incident intensity provides an additional operating degree of freedom for tuning radial localization and circulation, with the associated photothermal load defining the corresponding operating trade-off.

The Brownian-dynamics results in Figure S20 were evaluated at a fixed bath temperature of 298.15 K to isolate the effect of optical-force amplitude; the calculated photothermal temperature rise was not fed back into the viscosity or diffusion coefficient.

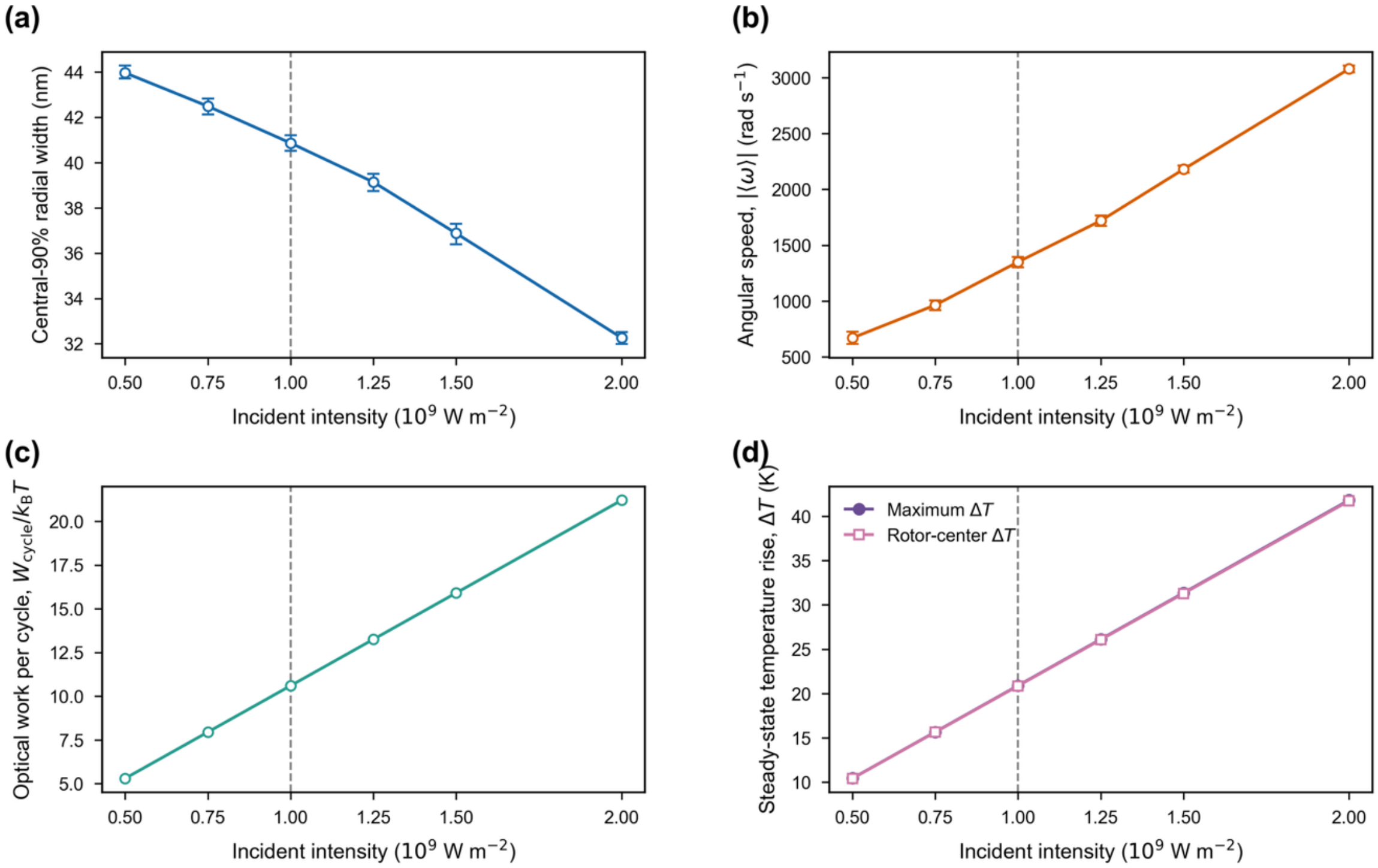

**Figure S20. Incident-intensity dependence of the orbital dynamics of C2. (a)** Central-90% radial width, **(b)** ensemble-averaged angular speed, **(c)** optical work per cycle normalized by $k_BT$, and **(d)** maximum and rotor-center steady-state temperature rises as functions of incident intensity. Brownian-dynamics results were obtained from 32 paired trajectories per intensity using the fixed-bath FDT-consistent model at 298.15 K. The dashed line marks the $1 \times 10^9$ W m$^{-2}$ reference intensity.